\documentclass[structabstract]{aa}
\usepackage{graphicx}
\usepackage{amsmath}
\usepackage{amssymb}

\def\be{\begin{equation}}
\def\ee{\end{equation}}
\def\bea{\begin{eqnarray}}
\def\eea{\end{eqnarray}}

\begin{document}

\title{Simulations of the Hyades}


\author{
A. Ernst\inst{1}
\and
A. Just\inst{1}
\and
P. Berczik\inst{1,3,4,5}
\and
C. Olczak\inst{1,2,3,4}
}

\institute{
Astronomisches Rechen-Institut, Zentrum f\"ur Astronomie der Universit\"at Heidelberg, 
M\"onchhofstrasse 12-14, D-69120 Heidelberg, Germany \\
e-mail: {\tt aernst@ari.uni-heidelberg.de}
\and
Max-Planck-Institut f{\"u}r Astronomie, K{\"o}nigstuhl 17, D-69117
Heidelberg, Germany
\and
National Astronomical Observatories of China, Chinese Academy of Sciences,
Datun Lu 20A, Chaoyang District, Beijing 100012, China 
\and
The Kavli Institute for Astronomy and Astrophysics at Peking University, 
Yi He Yuan Lu 5, Hai Dian Qu, Beijing 100871, China
\and
Main Astronomical Observatory, National Academy of
Sciences of Ukraine, Akademika Zabolotnoho 27, 03680 Kyiv, Ukraine 
}

\date{Received ... Accepted ...}

\abstract{Using recent observational data we present $N$-body 
simulations of the Hyades open cluster.}{We make an attempt to
determine the initial conditions of the Hyades cluster at the time of its formation 
to reproduce the present-day cumulative mass profile, stellar mass and luminosity 
function (LF).}{We performed
direct $N$-body simulations of the Hyades in an analytic Milky Way potential.
They account for stellar evolution and include primordial binaries in a few models.
Furthermore, we applied a Kroupa initial mass function and used extensive 
ensemble-averaging.}{We find that
evolved single-star King initial models with King parameters $W_0 = 6- 9$ and initial particle
numbers $N_0 = 3000$ provide good fits to the present-day observational 
cumulative mass profile within the Jacobi radius. The best-fit King model has an initial mass of
$1721\ M_\odot$ and an average mass loss rate of $-2.2 \ M_\odot/\mathrm{Myr}$.
The K-band LFs of models and observations
show reasonable agreement. Mass segregation is detected in both
observations and models.
If 33\% primordial binaries are included, the initial particle 
number is reduced by 5\% as compared to the model without primordial binaries.}{The 
present-day properties of the Hyades can be reproduced well by a standard King or Plummer 
initial model when choosing appropriate
initial conditions. The degeneracy of good-fitting models can be  quite high 
due to the large dimension of the parameter space. 
More simulations with different Roche-lobe filling factors and primordial binary fractions 
are required to explore this degeneracy in more detail.}



\authorrunning{A. Ernst et al.}
\titlerunning{Simulations of the Hyades}

\maketitle

\section{Introduction}

The Hyades (``rain stars'') 
are one of the best-known open star clusters on the northern sky. 
They are located in the constellation Taurus,
close (on the celestial sphere) to the Pleiades or ``Seven Sisters'', 
the most prominent open star cluster on the northern night sky. 
The Hyades and the Pleiades form the ``golden gate of the ecliptic'', 
i.e. the ecliptic separates the two clusters on the celestial sphere.
The Hyades cluster is markedly older than the Pleiades cluster.
It has an age of $625 \pm 50$ Myr derived from the helium abundance
in combination with isochrone modeling which includes convective 
overshooting (Perryman et al. 1998). The shape of the Hyades 
Hertzsprung-Russell (HR) diagram with a narrow and
well-defined main sequence has been recently studied by von Leeuwen (2009).
Previous $N$-body simulations of the Hyades have been performed
by Portegies Zwart et al. (2001), Madsen (2003; see also references therein) and 
Chumak, Rastorguev \& Aarseth (2005).
These studies are in many respects similar to the present study (although special
emphasis is placed on different aspects of the modeling of the Hyades).
The study by Portegies Zwart et al. (2001) differs from the present study by the 
fact that they did not include a kick for the white dwarfs but argued
that they could be hidden in binary systems with considerably
brighter companion stars.
Madsen (2003) concentrated on investigating the accuracy of astrometric 
radial velocities and compared the HR diagrams and internal velocity dipersions 
of observations and models. On the other hand, 
Chumak, Rastorguev \& Aarseth (2005) studied the Hyades orbit and provided
first models of the tidal tails of the Hyades.

The starting point for the present work is an observational data
file (R{\"o}ser et al. 2011) from which masses, positions,
and velocities for 724 probable member stars of the Hyades can be derived. 
 The PPMXL catalog (R\"oser, Demleitner \& Schilbach 2010) contains positions 
 and proper motions of the Hyades members 
down to $0.116 \ M_{\odot}$.\footnote{Note that there is a recent study (Goldman et al., in prep.) based on
PanSTARRS1 which goes down to even lower masses.}
The membership has been determined with the
convergent-point method. Given the right ascension, declination, and the respective
proper motions, the convergent-point method predicts a heliocentric distance 
(secular parallax) and radial velocity for each candidate member.
Thus the full 6D phase space information is given.
R{\"o}ser et al. determined the individual masses 
 from the mass-to-luminosity relation by
Pinsonneault et al. (2004), from Dartmouth isochrones
(Dotter et al. 1998), and from BCAH isochrones (Baraffe et al. 1998).
Binarity is known only for a minor portion of the sample (R{\"o}ser et al. 2011).

Table \ref{tab:obs-par} gives a few parameters of the Hyades cluster
as determined from the observational data file. We determined the 
tidal (or Jacobi) radius (King 1962) iteratively from the observational data
according to the procedure described in Section 4 of Ernst et al. (2010) 
under the assumption that the solar radius is at $R_0 = 8$ kpc (values for $R_0=8.5$ kpc
are given in brackets) and that $\beta=\kappa/\Omega = 1.37$, where $\kappa$ and
$\Omega$ are the epicyclic and circular frequencies of a near-circular
orbit at the solar radius. The incompleteness in the stellar mass function
sets in somewhere between $0.25$ and $0.17 \, M_{\odot}$ (R\"oser et al. 2011).


\begin{table}
\caption{Hyades parameters as determined from the observational data file.}
\label{tab:obs-par}
\begin{center}
\begin{tabular}{lc}
\hline\noalign{\smallskip}
Quantity & Value \\
\noalign{\smallskip}
\hline
\noalign{\smallskip}
  Lowest mass $m_l$ & $0.116 \ M_{\odot}$ \\
 Highest mass $m_h$ & $2.6 \ M_{\odot}$ \\ 
Maximum radius $r_{\rm max}$ & 30 pc \\
\ \ \ Number of stars $N(30 \ pc)$ & $724$ \\
\ \ \ Total mass $M(30 \ pc)$ & $469 \ M_{\odot}$ \\
Jacobi radius $r_J$ & $8.6$ pc ($9.0$ pc) \\
\ \ \ Number of stars $N(r_J)$ & 354 (359) \\  
\ \ \ Enclosed mass $M(r_J)$ & $272 \ M_{\odot} (275 \ M_{\odot})$ \\ 
Core radius $r_c$ & 3 pc \\
\ \ \ Number of stars $N(3 \ pc)$ & 97 \\  
\ \ \ Enclosed mass $M(3 \ pc)$ & $99 \ M_{\odot}$ \\ 
Completeness limit & $0.25 \dots 0.17 \ M_\odot$ \\
\hline
\end{tabular}
\end{center}
\end{table}


This paper is organized as follows: Section \ref{sec:numericalmethod}
describes the numerical methods we use. Section \ref{sec:orbit} describes
our integration of the Hyades orbit. In Section \ref{sec:parameterspace}
we discuss our parameter space in detail. Section \ref{sec:results}
summarizes the results. Section \ref{sec:discussion}
contains the discussion and conclusions.

\label{sec:introduction}

\section{Numerical method}

\label{sec:numericalmethod}

\begin{table}
\caption{The list of galaxy component parameters.}
\label{tab:gal-par}
\begin{center}
\begin{tabular}{lcrr}
\hline\noalign{\smallskip}
Component & M [M$_\odot$] & $a~[{\rm kpc}]$ & $b~[{\rm kpc}]$ \\
\noalign{\smallskip}
\hline
\noalign{\smallskip}
 Bulge & $1.4 \times 10^{10}$ & 0.0 &  0.3 \\
 Disk  & $9.0 \times 10^{10}$ & 3.3 &  0.3 \\
 Halo  & $7.0 \times 10^{11}$ & 0.0 & 25.0 \\ 
\hline
\end{tabular}
\end{center}
\end{table}

\begin{figure}[t]
\includegraphics[angle=-90,width=0.45\textwidth]{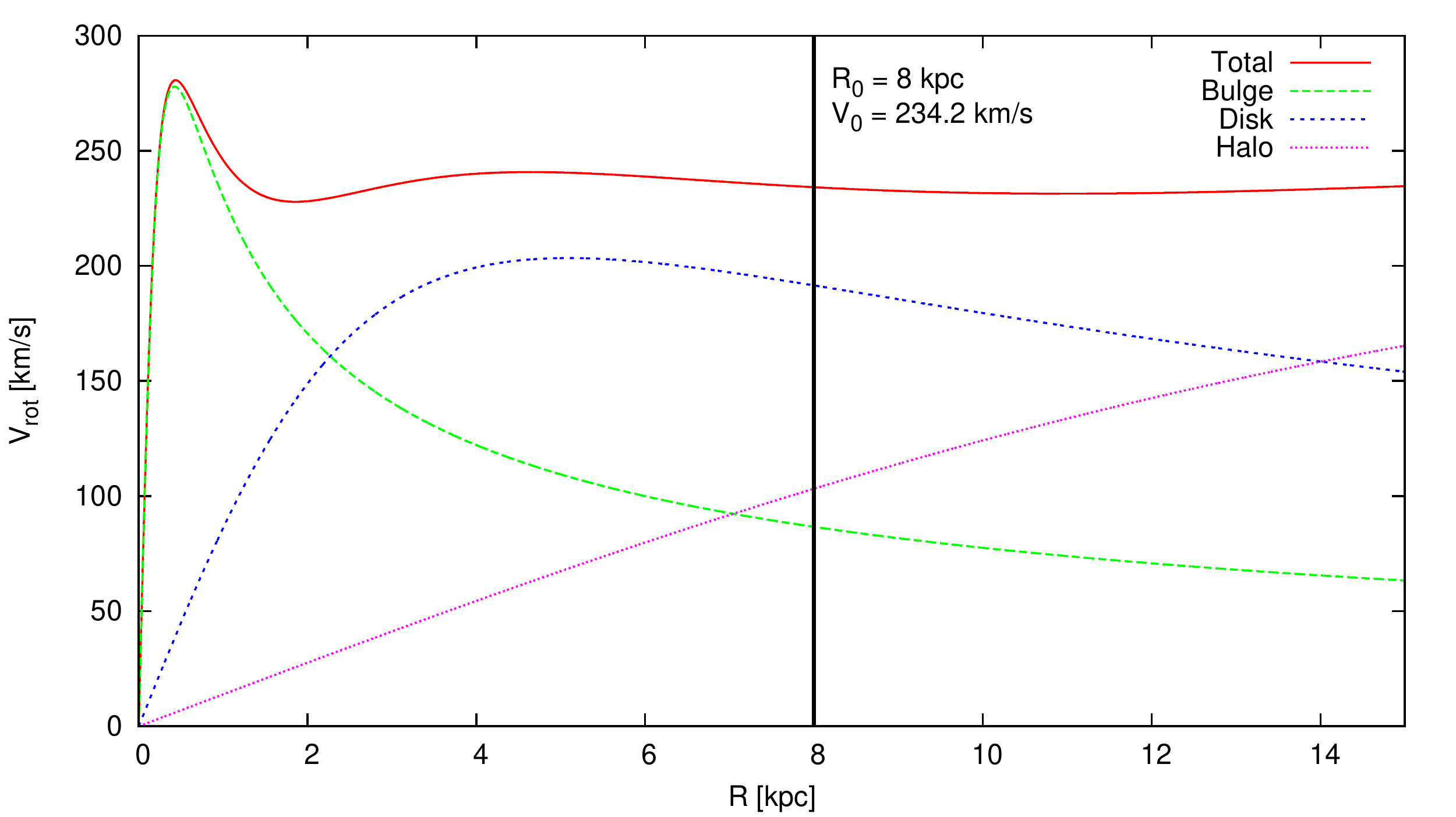} 
\caption{Rotation curve (at $z=0$) of the three-component Plummer-Kuzmin 
model of the Milky Way.} 
\label{fig:vrot}
\end{figure}

We solve the $N$-body problem of the Hyades in an analytic background potential
of the Milky Way. 
For the background Milky Way potential, we use an axisymmetric three-component model, 
where the bulge, disk, and halo are described by Plummer-Kuzmin models 
(Miyamoto \& Nagai 1975) with the potential

\be
\Phi(R,z) = - \frac{ GM }{ \sqrt{R^2 + (a + \sqrt{b^2 + z^2} )^2} }. \label{eq:eq-gal}
\ee

\noindent
The parameters $a, b$, and $M$ of the Milky Way model are given in Table \ref{tab:gal-par}
for the three components. In the limit $a\rightarrow0$ this model reduces to a Plummer model
(Plummer 1911). On the other hand, in the limit $b\rightarrow 0$ the model reduces to
a Kuzmin model (Kuzmin 1956).

Figure \ref{fig:vrot} shows the rotation curve of the three-component model of the Milky Way.
As in our previous works, the parameters of the three-component model
are chosen such that the rotation curve matches that of the Milky Way (Dauphole \& Colin 1995). 
At the solar radius $R_0=8.0$ kpc, which we have assumed in this study, 
the value of the circular velocity is 
$V_0=234.2$ km/s. The values of Oort's constants $A$ and $B$ are consistent 
with the observed values $(A,B) = (14.5 \pm 0.8, -13.0 \pm 1.1)$ km/s/kpc derived by
Piskunov et al. (2006).

\begin{figure}
\includegraphics[width=0.5\textwidth]{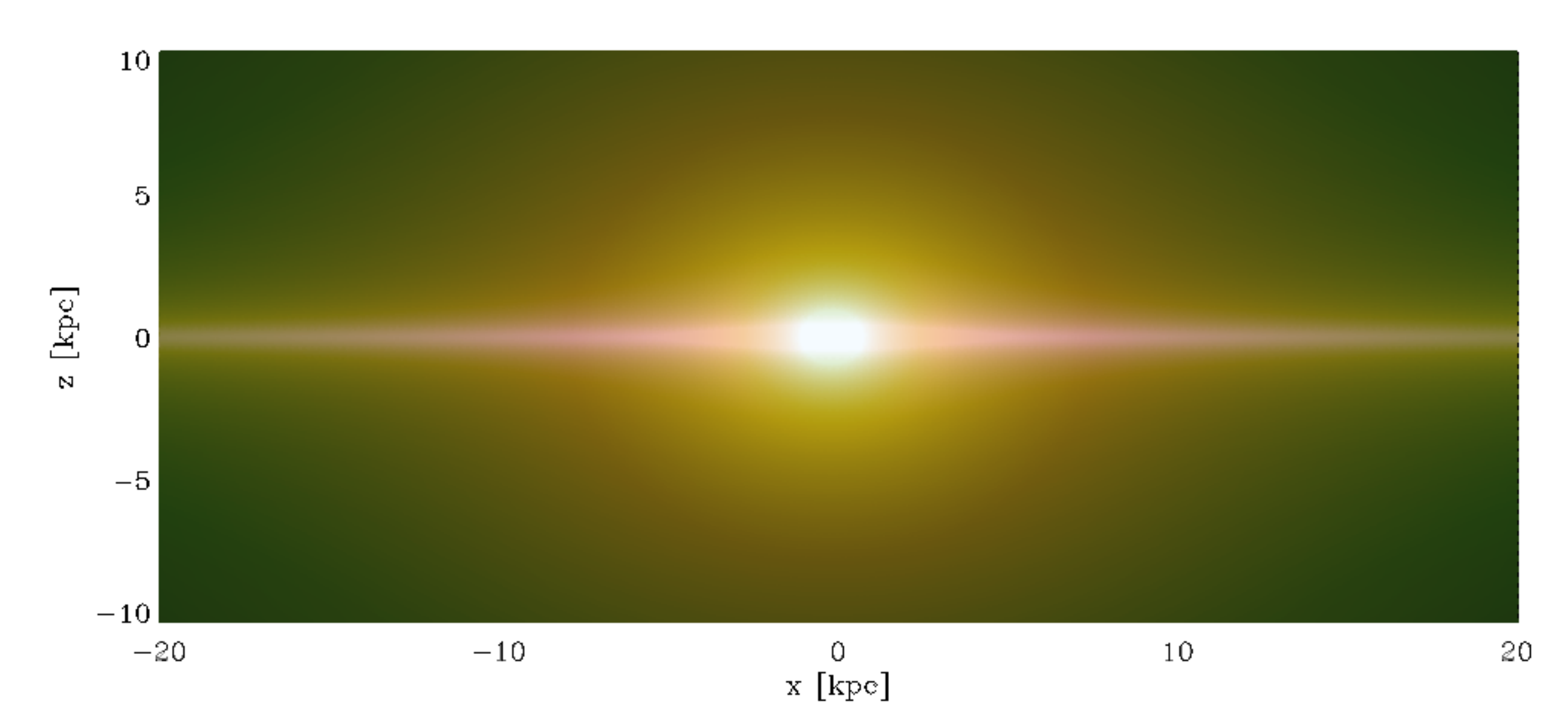} 
\caption{``Artistic'' RGBA image of the (logarithmic) density structure of the three-component 
Plummer-Kuzmin model.
The three components are overblended with an alpha channel with 90\% 
transparency. The halo is plotted in blue, the disk in green, and the bulge in red.} 
\label{fig:dens}
\end{figure}

Figure \ref{fig:dens} shows an ``artistic'' RGBA image of the (logarithmic) density structure of the 
three-component Plummer-Kuzmin ¥el.
The three components are overblended with an alpha channel with 90\% 
transparency. The halo is plotted in blue, the disk in green, and the bulge in red.

For the solution of the $N$-body problem in the analytic background potential
of the Milky Way, the new direct $N$-body code {\sc nbody6tidgpu} (see Appendix A for a 
description) is used. Since this program is based on the code {\sc nbody6} by Aarseth 
(see Aarseth 1999, 2003), it is possible to follow stellar evolution within the 
Hertzsprung-Russell diagram. The implementation 
of stellar evolution for individual stars of the $N$-body system is based on analytic formulae
from which radius, luminosity, and stellar type can be derived as functions of the initial mass and age
(Aarseth 2003; for the stellar evolution recipes see Hurley, Pols \& Tout 2000
and Hurley, Tout, Aarseth \& Pols 2001). 
In particular, this approach includes the formation of stellar remnants like
white dwarfs (WDs), neutron stars (NSs) and black holes (BHs). In addition to modeling standard evolutionary mass loss 
by stellar winds (which is implemented in many $N$-body codes available today), 
Aarseth's code and its variants include routines for treating large instantaneous 
mass loss due to special events like the occurrence of supernovae and the formation of planetary 
nebulae. (It is not very easy to correct energy, all forces, and first derivatives
within the $N$-body system for large instantaneous mass loss of one particle
in the case of, say, a supernova.) Thus it is possible to model 
a star cluster realistically with {\sc nbody6} (and its variants) even in the first 55 Myrs of 
evolution in which the supernova events occur (e.g. Hansen \& Kawaler 1994, their 
Equation (1.88)).

The code {\sc nbody6} and its variants optionally apply velocity kicks to stellar 
remnants when a supernova occurs or a planetary nebula forms. The random 
kick velocities for supernova kicks are drawn from a Maxwellian distribution 
with a 1D velocity dispersion of $\approx 190$ km/s corresponding
to a mean of $\approx 300$ km/s (Hansen \& Phinney 1997).
For all WDs, the 1D dispersion of the Maxwellian is taken to 
be $5$ km/s which corresponds to a mean of $\approx 8$ km/s 
(e.g. Fellhauer et al. 2003 for a lower limit on that dispersion).
We stress that the final number of WDs in our models depends
critically on this dispersion.
The escape velocity from the center of the cluster
is $\approx 6-7$ km/s for an open cluster with a mass of $2000 \ M_\odot$ 
and a half-mass radius of $3.5$ pc, assuming that it has a Plummer
profile (Plummer 1911). 

Finally, we note that in {\sc nbody6tidgpu}, all stars are kept in the 
simulation forever, i.e. there is no optional removal of escapers as in {\sc nbody6}
and its other variants. The reason is that we would like to follow the
stellar orbits in the tidal tails and investigate their properties.

\section{Orbit}

\label{sec:orbit}

\begin{figure}[b]
\centering
\includegraphics[angle=-90,width=0.5\textwidth]{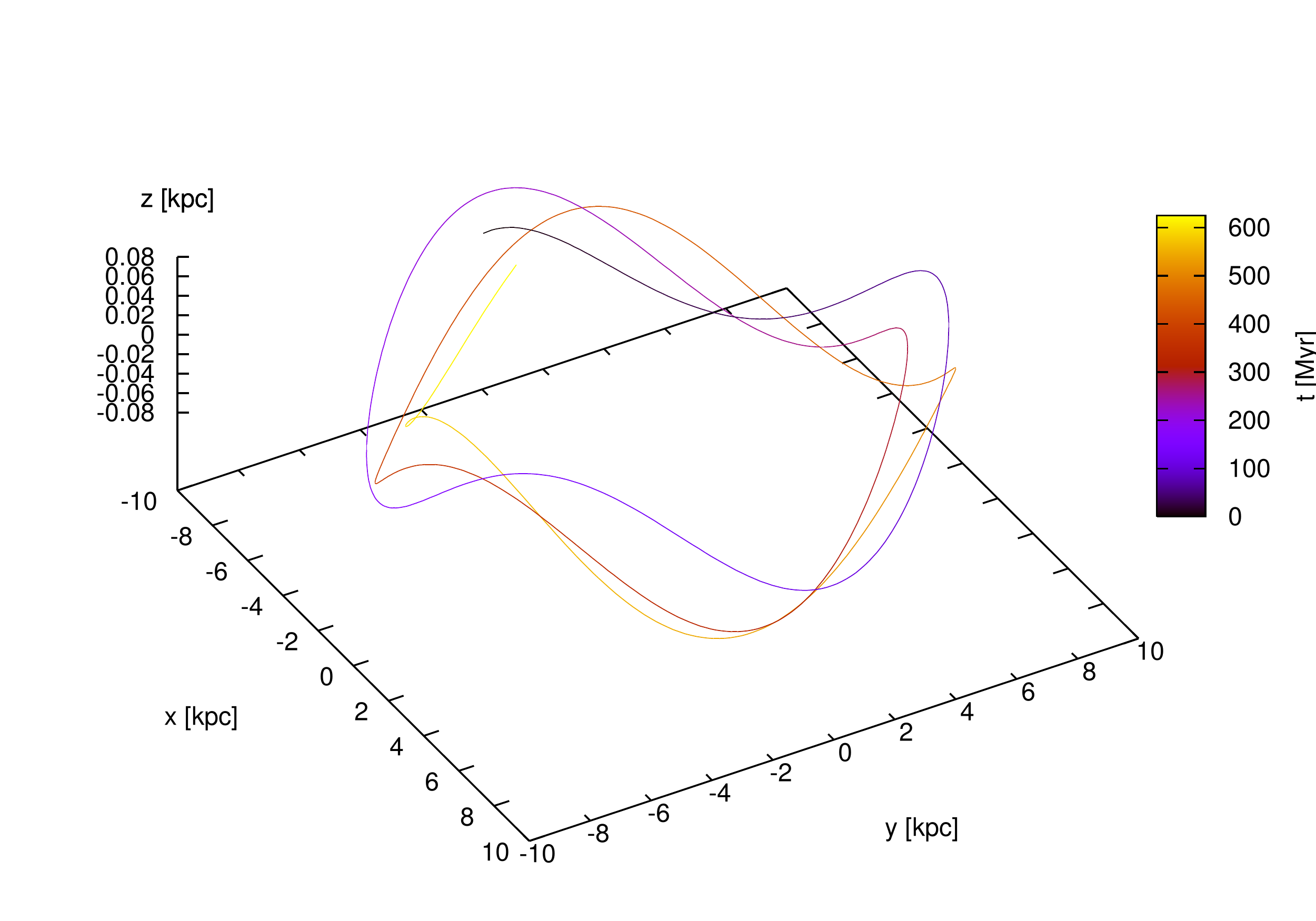} 
\caption{3D orbit of the Hyades in the three-component Plummer-Kuzmin
model of the Milky Way. The integration time is $625$ Myrs. 
} 
\label{fig:ernst_comp3}
\end{figure}

We integrated the center of mass orbit of the Hyades backwards in time in the 
analytic Milky Way potential (i.e. the three-component Plummer-Kuzmin model
described in section \ref{sec:numericalmethod}).
We neglected encounters with giant molecular clouds and the effect
of spiral arm passages and disk shocking. 
The integration time (625 Myr) was equal to the most probable age 
of the Hyades (Perryman et al. 1998). The full 3D orbit is shown in
Figure \ref{fig:ernst_comp3}.
The orbit integration with a simple integrator yielded the initial
position of the Hyades at the time of its formation.
From the initial position of the Hyades
we ran full $N$-body models up to the present time.

The present-day position of the Hyades was taken to be

\bea
x_0 &=& -44.26 \ \mathrm{pc} - 8000.00 \ \mathrm{pc}, \label{eq:x1} \\
y_0 &=& 0.31 \ \mathrm{pc}, \label{eq:x2} \\
z_0 &=& -16.89 \ \mathrm{pc} + 20.00 \ \mathrm{pc} \label{eq:x3}. 
\eea

\noindent
The first terms in Eqs. (\ref{eq:x1}) - (\ref{eq:x3}) correspond to the
center of mass of the observed Hyades spatial distribution with respect to 
the position of the Sun.
The second terms in Eqs. (\ref{eq:x1}) and (\ref{eq:x3}) represent
the solar position in the Milky Way. We used $(R_0,z_0) = (8.00, 0.02)$ kpc 
(consistent with Piskunov et al. 2006).

The present-day velocity was given by

\bea
v_{x0} &=& 41.18 \ \mathrm{km \ s}^{-1} - 11.10 \ \mathrm{km \ s}^{-1}, \label{eq:v1} \\
v_{y0} &=& 19.04 \ \mathrm{km \ s}^{-1} - 12.24 \ \mathrm{km \ s}^{-1} + 234.20 \ \mathrm{km \ s}^{-1}, \label{eq:v2}\\
v_{z0} &=& 1.27 \ \mathrm{km \ s}^{-1} - 7.25 \ \mathrm{km \ s}^{-1} \label{eq:v3}.
\eea

\noindent
The first terms in Eqs. (\ref{eq:v1}) - (\ref{eq:v3}) represent the
velocity of the center of mass of the observed Hyades distribution with respect to the Sun.
The second terms in Eqs. (\ref{eq:v1}) - (\ref{eq:v3}) correspond
to the peculiar motion $(U,V,W)_\odot$ of the Sun with respect to the local 
standard of rest (LSR) according to Sch\"onrich et al. (2010).
The third term in Eq. (\ref{eq:v2}) is the orbital speed of the LSR 
at $R_0 = 8$ kpc in the three-component Plummer-Kuzmin model of Section
\ref{sec:numericalmethod}.

With these parameters, the Hyades orbit around the Galactic center
is oriented in a clockwise direction as seen from the Galactic north pole. The orbit
has a considerable vertical oscillation in the $z$-direction between
$\approx \pm 74$ pc. In the radial direction the orbit is mildly 
eccentric and oscillates between $7087$ pc and $8638$ kpc.
Roughly three orbits are completed in 625 Myrs.

\section{Parameter space}

\label{sec:parameterspace}

We were looking for the best-fitting King models (King 1966) and Plummer models
(Plummer 1911) to the observations of R{\"o}ser et al. after 625 Myrs of evolution with 
respect to the final cumulative mass profile $M(r)$ within $0 < r < 9$ pc (Jacobi radius).
We compare the cumulative mass profiles of 
models and the observational data and not the density profiles because 
the cumulative mass profiles are of a better quality with respect to noise.
Second, we note that the observational data of R{\"o}ser et al.
contains only 724 stars, but the observational present-day mass function (PDMF)
drops below the completeness limit, which lies between 0.25 and 0.17 $M_\odot$.
If we assume that the PDMF is shallow at the low-mass end according to Eq.
\ref{eq:IMF} below, $\approx 100$
stars would be missing in the low-mass bins above $0.116 \ M_\odot$
(see R{\"o}ser et al. 2011, their Figure 10). We took that into account
for the comparison of the particle numbers and total masses of observation
and model. 
 
All our models are initially Roche-lobe filling, i.e. 
the initial 99\% Lagrange radius $r_{99\%}$ was set to be equal to the 
tidal (or Jacobi) radius $r_J$. The 99\%
Lagrange radius is defined as the radius that contains 99\% of the 
cluster mass. The Jacobi radius is given by (King 1962)

\be
r_J = \left[ \frac{GM}{4\Omega^2-\kappa^2} \right]^{1/3} 
\ee

\noindent
where $G$, $M$, $\Omega$, and $\kappa$ are the
gravitational constant, the cluster mass, the circular, and the epicyclic 
frequencies of a near-circular orbit, respectively. 

We defined the initial mass function (IMF) as the number of stars per unit 
logarithmic mass interval. It is given by

\be
\frac{dN}{d\ln m} = \left\{ \begin{array}{cl}
c_1 \, m^{+0.7} & \ \ \ \ \ 0.01 \leq m/M_{\odot} < 0.08  \\
c_2 \, m^{-0.3} & \ \ \ \ \ 0.08 \leq m/M_{\odot} < 0.50  \\
c_3 \, m^{-1.3} & \ \ \ \ \ 0.50 \leq m/M_{\odot} < 1.00 \\
c_4 \, m^{-1.3} & \ \ \ \ \ 1.00 \leq m/M_{\odot}  \\
\end{array}\right. \label{eq:IMF}
\ee

\noindent
where $c_1 - c_4$ are four constants (determined
by the normalization and three conditions for the transitions
between the mass regimes). The exponents in our Eq. (\ref{eq:IMF}) are the 
exponents $\alpha_0 - \alpha_3$ given in Kroupa (2001), his Eq. (2), adjusted for
logarithmic mass bins. Thus our IMF is a Kroupa (2001) IMF.
We forced the lowest and the highest mass to take on the values 
$m_{l}=0.08 \ M_\odot$ and $m_{h}=100.00 \ M_\odot$, respectively.
Thus the first power law in (\ref{eq:IMF})
for the brown dwarf regime is obsolete.
In {\sc nbody6tidgpu}, the individual masses were randomly drawn from 
the above distribution (\ref{eq:IMF}) using a routine by Kroupa with 
a correction by Weidner.
 
For all models we switched on stellar evolution in {\sc nbody6tidgpu}. 
For the metallicity we used the observed value for the Hyades $Z = 0.024$ 
(Perryman et al. 1998).

More than 400 individual models have been computed. The run time for
one model on a dual-Xeon 3.2 GHz with a GeForce 8800 GTS 512 graphics
card was about 45 minutes.

In a first step, a grid of 30 King models (King 1966) in the parameter space 
($W_0$, $N$), where $W_0$ is the King parameter and $N$ the initial particle 
number, was integrated for an overview of 
the evolution of different models.
We used for the King parameters $W_0=3,4,5,6,7,8,9,10,11,12$ and particle 
numbers $N=3000,4000,5000$. Thus the grid cell size in the parameter space
is given by $(\Delta W_0, \Delta N)=(1,1000)$.
We did not include primordial binaries in these models.

Using this first grid, we were able to notice trends in the choices of 
$N$ and $W_0$. The variation in $W_0$ turned out to be weak.
Also, the grid was too coarse to reach the desired accuracy in $N$. 

In the second step, we therefore continued with an iterative search in the 
parameter space ($W_0, N$) for the best-fitting models to the observations 
without primordial binaries. 
For this purpose we refined our first grid in the parameter space with 
respect to $N$ to a grid cell length of $\Delta N=125$ particles. 
On the other hand, it turned out to be useful to coarsen the grid with 
respect to $W_0$ to a grid cell height of $\Delta W_0=3$, i.e. we restricted the 
investigation to the models with $W_0=3, 6, 9, 12$. Thus the grid cell size of the second
grid in the parameter space was given by $(\Delta W_0, \Delta N)=(3,125)$.

Our criterion for the best fit is that the inner cumulative
mass profiles within $r = 9$ pc (Jacobi radius) of observations (R\"oser et al. 2011,
their Figure 6) and models (ensemble mean, see below) show the best agreement.

We did not run all models in this second grid. Rather, 
we started from the best-fitting model in the first grid for a given $W_0$ 
and varied the particle number $N$ in steps of $\Delta N$=125 particles in order
to find the best-fitting model in the second grid. 

At the same time, we performed $n=15$ runs with different random number seeds
for each initial model in the second grid. From these 15 runs we obtained 
physical quantities by ensemble averaging. For example, we compared the
observations with the mean cumulative mass profile of 15 models to find
the best fit acording to the criterion mentioned above. 
The random number generator is invoked mainly for the following purposes:

\begin{itemize}
\item Drawing of initial masses, positions, and velocities from
the given probability distributions (Kroupa 2001, King 1966);
\item Assignment of kick velocities for WDs, NSs,
and BHs;
\item Assignment of perihelion, node, and inclination of primordial binaries.
\end{itemize}

We notice that we did not simply ensemble-average the initial model from
15 initial models, which differed by their random number seeds.
In particular, we did not perform only one single run from such an ensemble-averaged 
initial model but really performed 15 individual runs, because

\begin{itemize} 
\item we are looking for standard deviations in the physical
quantities,
\item the modelling of the population of stellar remnants 
should be as realistic as possible.
\end{itemize}

\noindent
In order to verify that we found indeed the best-fitting ensembles, we 
looked at the cumulative mass profile of at least one of the neighboring 
ensembles with respect to $N$.

The ensembles in the second parameter grid are named, for example, en3000W6 
($N_0 = 3000, W_0=6$), i.e. the letters ``en'' for ``ensemble''
are followed by the initial particle number $N_0$ and the King parameter $W_0$. 



We also looked for the best-fitting ensemble of single-star Plummer models (Plummer 1911)
with respect to the inner cumulative mass profile (within $r=9$ pc)  also using
steps of $\Delta N = 125$. The choice of $r_{99\%}/r_J$ fixes the Plummer radius,
while the King models still have the free parameter $W_0$. The Plummer model runs 
are discussed in Sect. \ref{sec:primordialbinaries}.

We ran the best-fitting ensemble of single-star Plummer models with primordial binaries 
in order to see their effect on the evolution.
For models with primordial binaries we defined the primordial 
binary fraction $f_b$ and the fraction of stars in binaries $f_{sb}$ as

\be
f_b = \frac{N_b}{N_s + N_b}, \ \ \ \ \ f_{sb} = \frac{2 N_b}{N_s + 2 N_b} \label{eq:fb}
\ee

\noindent
where $N_s$ and $N_b$ are the initial numbers of binaries and 
single stars, respectively (e.g. Kroupa 1995, Jahrei{\ss} \& Wielen 2000, 
Heggie, Trenti \& Hut 2006).
The period-generating function for the primordial binaries 
was given by (Kroupa 1995, Eq. 11b)

\be
\log_{10} P(X) = \log_{10}P_{\rm min} + \left\{\delta\left[\exp(2X/\eta)-1\right]\right\}^{1/2}
\ee

\noindent with a uniform random variate $X \in [0,1]$.
The parameters are $\eta = 2.5$, $\delta = 45.0$, and $P_{\rm min} = 5$ days.
For the binary eccentricities a thermal distribution
was adopted with the eccentricity-generating function
$e^2 = X$ where $e$ is the eccentricity and $X \in [0,1]$ 
a uniform random variate. We did not assume any correlation between the 
masses of the binary components (e.g. Eggleton, Fitchett \& Tout 1989).

\section{Results}

\label{sec:results}

For the statistical analysis, we applied the quantities

\be
\overline{x} = \frac{1}{n}\sum_{i=1}^n x_i, \ \ \
\sigma =\sqrt{\frac{1}{n-1}\sum_{i=1}^n \left(x_i-\overline{x}\right)^2}, 
\ \ \ \sigma_m = \frac{\sigma}{\sqrt{n}},
\ee

\noindent
where $n=15$ is the number of models in an ensemble, $x_i$  are the individual measurements,
$\overline{x}$ is their mean,
$\sigma$ the standard deviation (i.e. the mean error of a measurement in a single model),
and $\sigma_m$ the standard error (i.e., the mean error of the mean).

\subsection{Cumulative mass profiles}

\label{sec:cumulativemassprofiles}

The contamination with field stars in the observational data
file is negligible for $r<9$ pc, 7.5\% at $9$ pc $< r < 18$ pc, 
and 30\% at 18 pc $< r < 30$ pc (with respect to the number of stars; 
R{\"o}ser et al. 2011), where $r$ is measured from the center of the cluster.

\begin{figure}[t]
\centering
\includegraphics[angle=-90,width=0.5\textwidth]{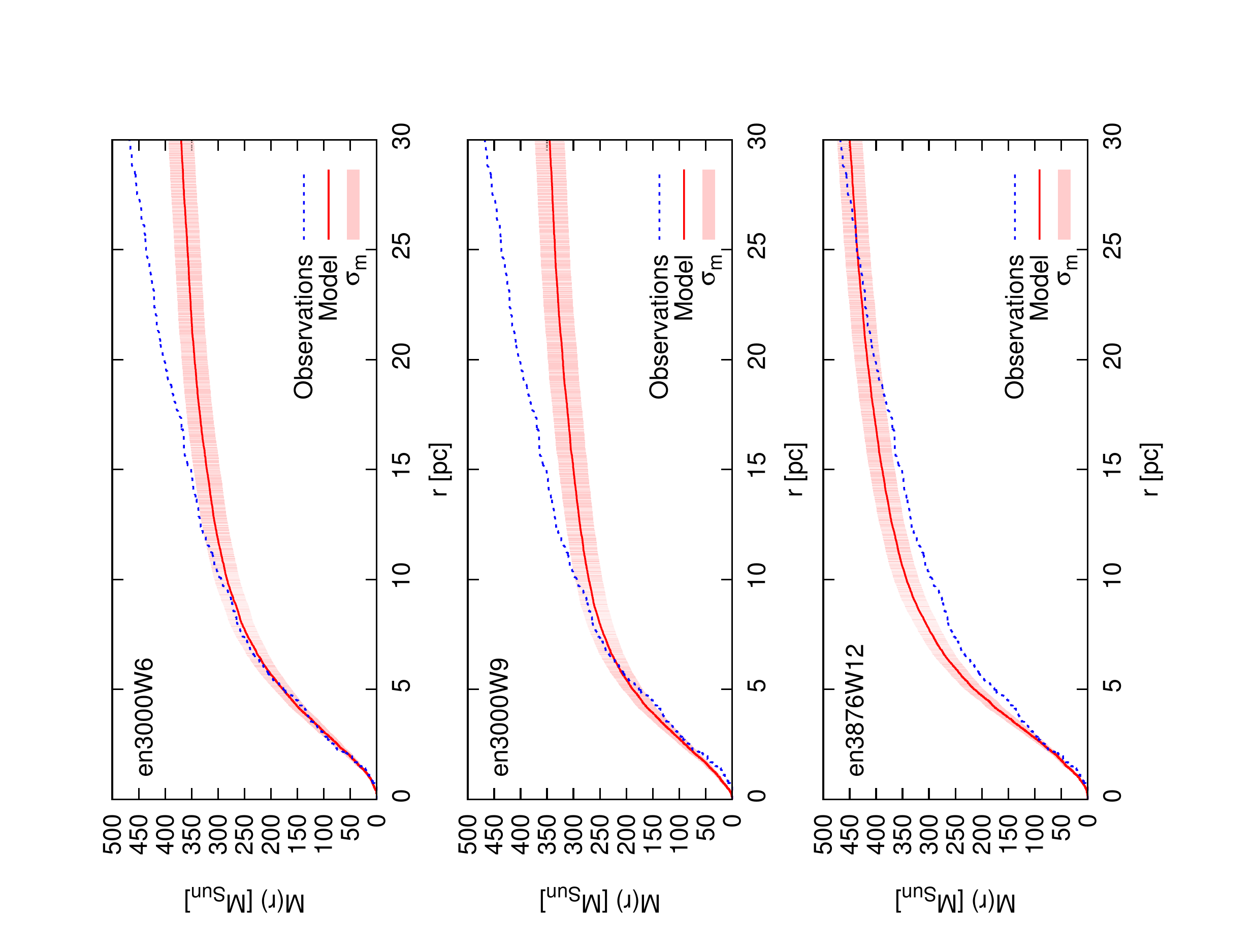} 
\caption{Comparison of the present-day cumulative mass profiles of observations
and models. The thick solid (red) lines in the middle of the models  
represent the mean, which is ensemble-averaged over 15 runs. 
The standard error is shown as a filled area.
The dashed (blue) line is calculated from the observational data.
Top: The ensemble en3000W6.
Middle: The ensemble en3000W9.
Bottom: The ensemble en3875W12.}
\label{fig:mrmean4}
\end{figure}

Three comparisons of the cumulative mass profiles are shown in Fig. \ref{fig:mrmean4}.
The top panel shows the comparison for the ensemble of models en3000W6 
($N_0 = 3000, W_0=6$). Shown is the mean cumulative mass profile of
the ensemble of 15 models, i.e., we averaged over the 15 single cumulative
mass profiles of the individual models.
Compared with the observations, 
the ensemble en3000W6 is the best-fitting of all our 
models according to the criterion stated in Section \ref{sec:parameterspace}.
Assuming that the model is perfect, the field star contamination is 
$|M_{\rm model}- M_{\rm obs}|/M_{\rm obs}=$ 
$2\% \ (3 \ \rm pc)$, $2\% \ (9 \ \rm pc)$, $12\% \ (18 \ \rm pc)$, $21\% \ (30 \ \rm pc)$, 
where $M_{\rm mod}$ and
$M_{\rm obs}$ are theoretical and observed cumulative masses.

The second-best fitting ensemble of King models (en3000W9; $N_0 = 3000, W_0=9$) 
is shown in the middle panel of Figure \ref{fig:mrmean4}. 
The ensemble mean of the cumulative mass profile also fits the inner part
of the observational profile fairly well.
The field star contamination here is $|M_{\rm model}- M_{\rm obs}|/M_{\rm obs}=$ 
$12\% \ (3 \ \rm pc)$,  $5\% \ (9 \ \rm pc)$, $18\% \ (18 \ \rm pc)$, $26\% \ (30 \ \rm pc)$.

The bottom panel shows the comparison of the cumulative mass profiles 
for the observations and the ensemble en3875W12 ($N_0 = 3875, W_0=12$). 
This model has been chosen because the agreement with the observations is 
very good in the core ($r<3$ pc) and for the final total mass contained within a 
radius of $r=30$ pc from the cluster center. 
In the range $3 \ \mathrm{pc} < r < 18 \ \mathrm{pc}$ pc, the fit is fairly bad, i.e. inconsistent
within $1\sigma_m$. A better fit of the core region cannot be obtained for $W_0=12$. 
For $W_0 = 3$ the models dissolve so fast that a good fit with respect to 
the inner cumulative mass profile or the final total mass has not been obtained 
at all. In Appendix \ref{sec:badfits} we show the best fits for the present-day
total mass within $r=30$ pc from the center, which, however, do not agree 
with the inner cumulative mass profile.

\subsection{Past evolution and present-day state}

\label{sec:pastevolution}

\begin{table}
\caption{Parameters of the best fitting ensemble of models en3000W6.}
\label{tab:en3000W6-par}
\begin{center}
\begin{tabular}{lll}
\hline\noalign{\smallskip}
\# & Quantity & Value\\
\noalign{\smallskip}
\hline
1 & Initial model (initial) & \\
\hline
\noalign{\smallskip}
& Particle number $N_0$ & $3000$ \\
& Total mass $\langle M_0 \rangle$ & $1721 \, M_{\odot}$ \\
& King parameter $W_0$ & $6$ \\
& Jacobi radius $r_J$ & 16.2 pc \\
\hline
2 & $r<30$ pc (final) & \\
\hline
& Particle number $\langle N_{\rm f}\rangle \pm \sigma_m$ & $736 \pm 50$ \\
& ``Observed'' part. numb. $\langle N_{\rm f} \rangle_{\rm obs} \pm \sigma_m$ & $625 \pm 42$ \\
& Lowest mass $m_l$ & $0.08 \, M_{\odot}$ \\
& Highest mass $m_h \pm \sigma_m$ & $2.68 \pm 0.01 \, M_{\odot}$ \\
& Mean mass $\langle m \rangle$ & $0.503 \, M_{\odot}$  \\ 
& Total mass $M_{\rm f} \pm \sigma_m$ & $369 \pm 25 \, M_{\odot}$ \\
\hline
3 & $r<9$ pc (final, inside Jacobi radius) & \\
\hline
& Particle number $\langle N_{\rm f}\rangle \pm \sigma_m$ & $476 \pm 43$ \\
& ``Observed'' part. numb. $\langle N_{\rm f} \rangle_{\rm obs} \pm \sigma_m$ & $414 \pm 37$ \\
& Mean mass $\langle m \rangle$ & $0.565 \, M_{\odot}$  \\ 
& Total mass $M_{\rm f} \pm \sigma_m$ & $269 \pm 25 \, M_{\odot}$ \\
\hline
4 & $r<3$ pc (final, inside core) & \\
\hline
& Particle number $\langle N_{\rm f}\rangle \pm \sigma_m$ & $124 \pm 14$ \\
& ``Observed'' part. numb. $\langle N_{\rm f} \rangle_{\rm obs} \pm \sigma_m$ & $112 \pm 12$ \\
& Mean mass $\langle m \rangle$ & $0.784 \, M_{\odot}$  \\ 
& Total mass $M_{\rm f} \pm \sigma_m$ & $97 \pm 11 \, M_{\odot}$ \\
\hline
\end{tabular}
\end{center}
\end{table}

\begin{figure}[t]
\centering
\includegraphics[angle=-90,width=0.45\textwidth]{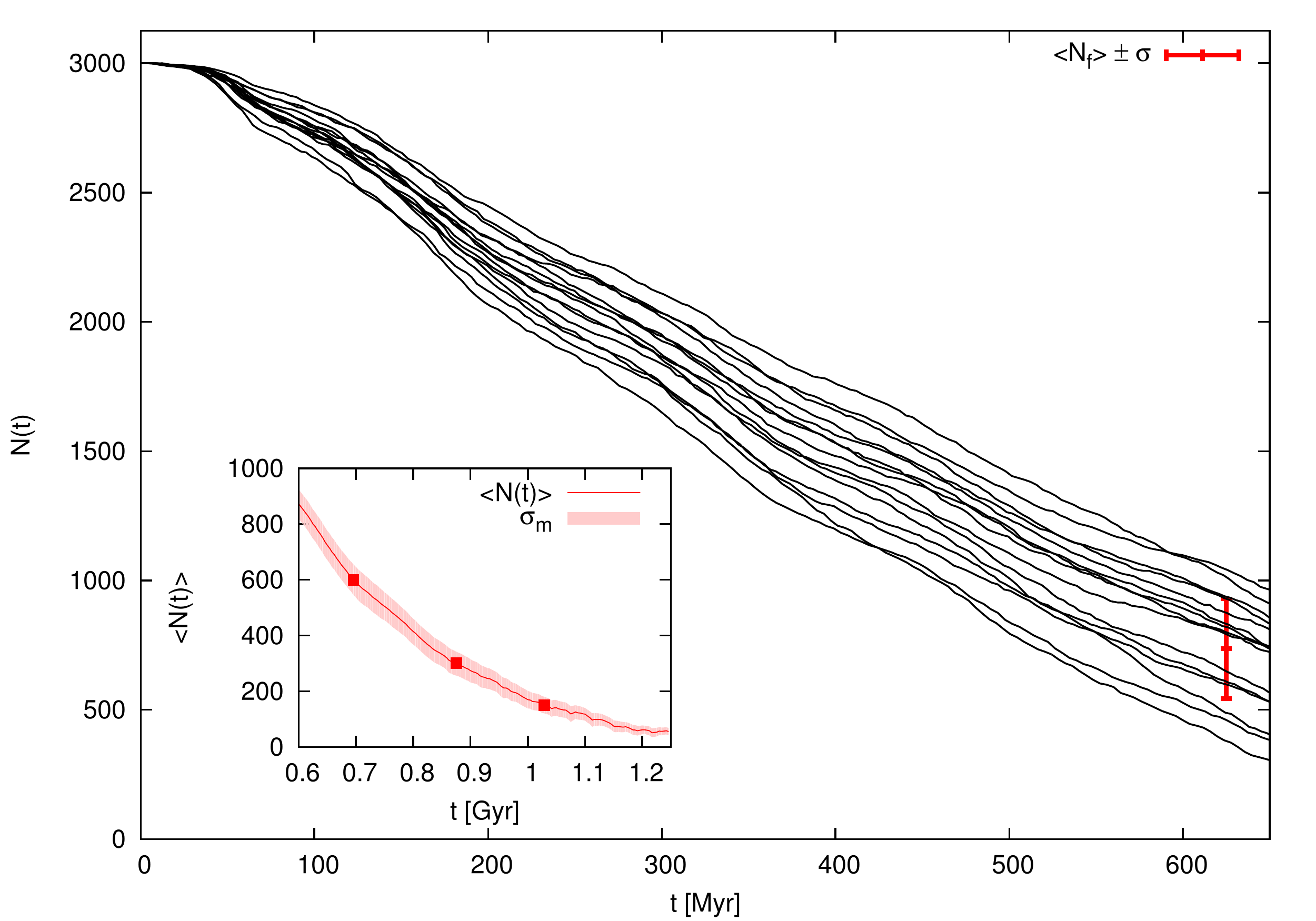} 
\caption{Particle number within $r=30$ pc from the cluster center 
as a function of time for all 15 runs of the ensemble en3000W6. 
One can see the non-Markovian process of star cluster dissolution.
The mean final particle number and the standard deviation for a single
run is shown as a (red) dot with vertical errorbars at $t=625$ Myrs. The future evolution
of the mean particle number within $r=30$ pc is shown in the inset. 
There the times when the cluster has reached
20, 10 and 5\% of its initial particle number are shown as (red) dots.} 
\label{fig:ntall}
\end{figure}

\begin{figure}[t]
\centering
\includegraphics[angle=-90,width=0.5\textwidth]{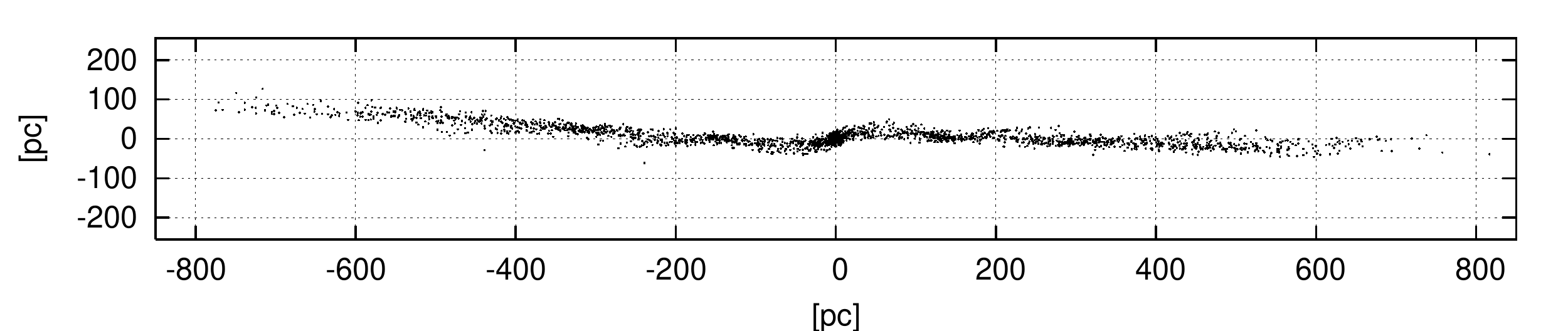} 
\caption{The present-day state of one run of the ensemble of models en3000W6 
after 625 Myrs of evolution. The tidal tails have reached a length 
of $800$ pc.} 
\label{fig:snapfin}
\end{figure}


\begin{table}
\caption{3D velocity dispersions at $t=625$ Myr for the ensemble en3000W6
for stars 30, 9 and 3 pc of the cluster center.}
\label{tab:vel-disp}
\begin{center}
\begin{tabular}{ll}
\hline\noalign{\smallskip}
Radius & $\langle\sigma_{\rm 3D}\rangle \pm \sigma_m$ [km/s] \\
\noalign{\smallskip}
\hline
\noalign{\smallskip}
 $r<30$ pc & $0.55 \pm 0.00$ \\
 $r<9$ pc  & $0.48 \pm 0.01$ \\
 $r<3$ pc  & $0.54 \pm 0.01$ \\ 
\hline
\end{tabular}
\end{center}
\end{table}


The best-fitting ensemble of models (en3000W6) for the inner cumulative 
mass profile is an ensemble average over 15 runs with different
random number seeds. The initial and final parameters of the ensemble en3000W6
are given in Table \ref{tab:en3000W6-par}. 
We note that (1) the final parameters are obtained after 625 Myrs of evolution, i.e. at the present time, (2) the final ``observed'' particle number is the number of stars with 
$m > 0.116 \ M_{\odot}$ and that (3) the highest mass is obtained by excluding BHs.

In the following discussion, we look at stars within distances of $r=30$ pc of the 
cluster center, within $r=9$ pc (Jacobi radius) and at stars within $r=3$ pc (core).
Fig. \ref{fig:ntall} shows the evolution of the particle number
within $r=30$ pc from the cluster center. It can be seen how the 15
cluster models of the ensemble en3000W6 dissolve as time proceeds.
The mean final particle number and the standard deviation for a single
run is shown with errorbars at $t=625$ Myrs.
The standard deviation is rather high due to the non-Markovian
evolution of the models, which produces rms scatter
between the individual models of an ensemble. 
The average particle loss rate until $t=625$ Myr is $dN/dt = -3.6/\mathrm{Myr}.$
The average mass loss rate until $t=625$ Myr is $dM/dt = -2.2 \ M_\odot/\mathrm{Myr}.$
The inset of Fig. \ref{fig:ntall} is explained in Sect. \ref{sec:futureevolution}.

Figure \ref{fig:snapfin} shows the present-day state of the cluster with
its tidal tails for one run of the ensemble en3000W6. The tidal tails have 
reached a length of 800 pc after 625 Myrs of evolution. However, the
kicked-out NSs and BHs (and a few WDs) 
hurried ahead along the cluster orbit during the evolution and reside 
at $t=625$ Myrs spread out along the orbit well beyond the tips of the tidal 
tails at 800 pc.


Table \ref{tab:vel-disp} shows the values of the 3D velocity dispersion 
$\sigma_{\rm 3D}$ for the ensemble en3000W6
within 30, 9, and 3 pc of the cluster center. The three values
are consistent with the fact that the velocity dispersion in the core is
higher than in the halo within the Jacobi radius and rises again outside
of the Jacobi radius. The values  can be compared with those given in R\"oser et al. (2011).

\subsubsection{Stellar mass functions}

\label{sec:stellarmassfunctions}

\begin{figure}[t]
\centering
\includegraphics[angle=-90,width=0.5\textwidth]{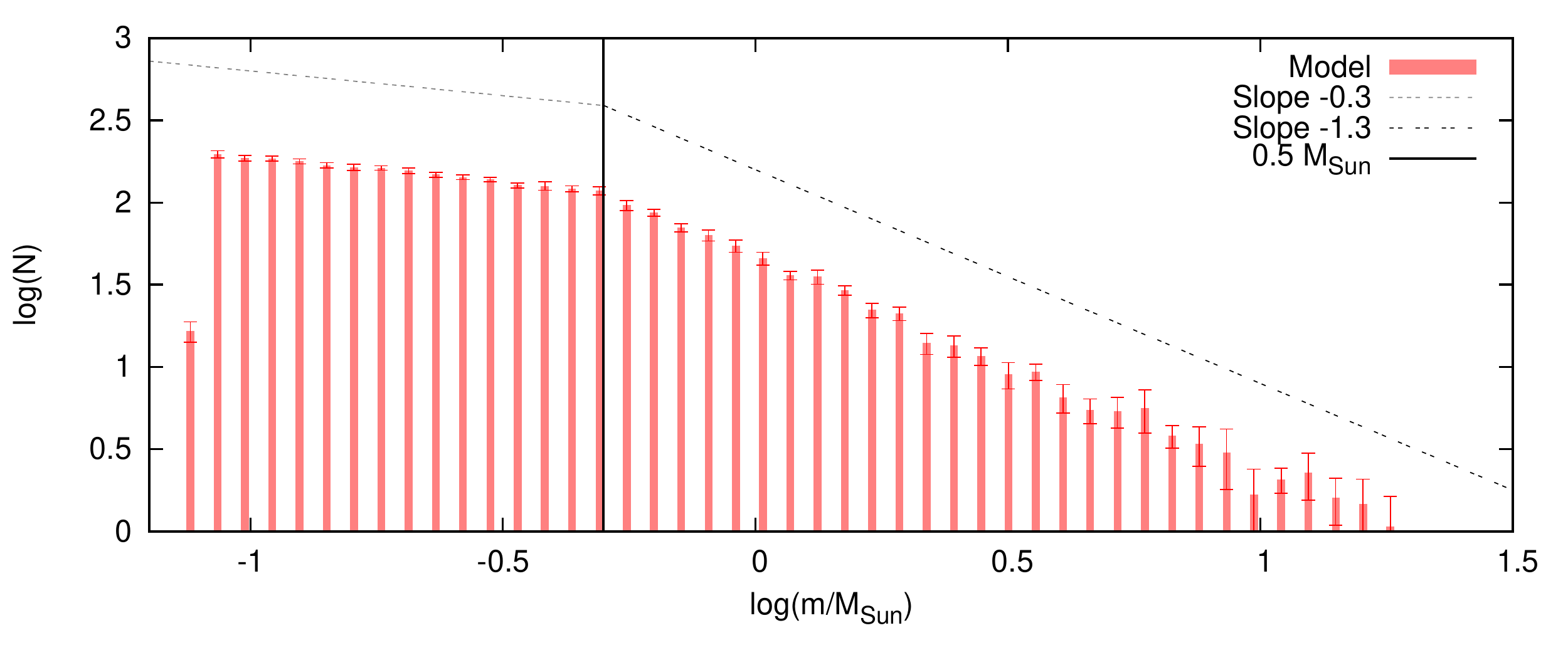} 
\caption{The average realization of the Kroupa (2001) IMF of the ensemble en3000W6 
with lowest mass $m_l = 0.08 \ M_\odot$, highest mass $m_h = 100.00  \ M_\odot$, 
and initial particle number $N_0=3000$. 
Shown is the mean value of $\log(N)$ from the ensemble averaging over 15 runs.
The $2\sigma_m$ standard errors are shown. The analytical
slopes and the transition at $0.5 \ M_\odot$ are also shown as lines.} 
\label{fig:histimf2}
\end{figure}

\begin{figure}[t]
\centering
\includegraphics[angle=-90,width=0.5\textwidth]{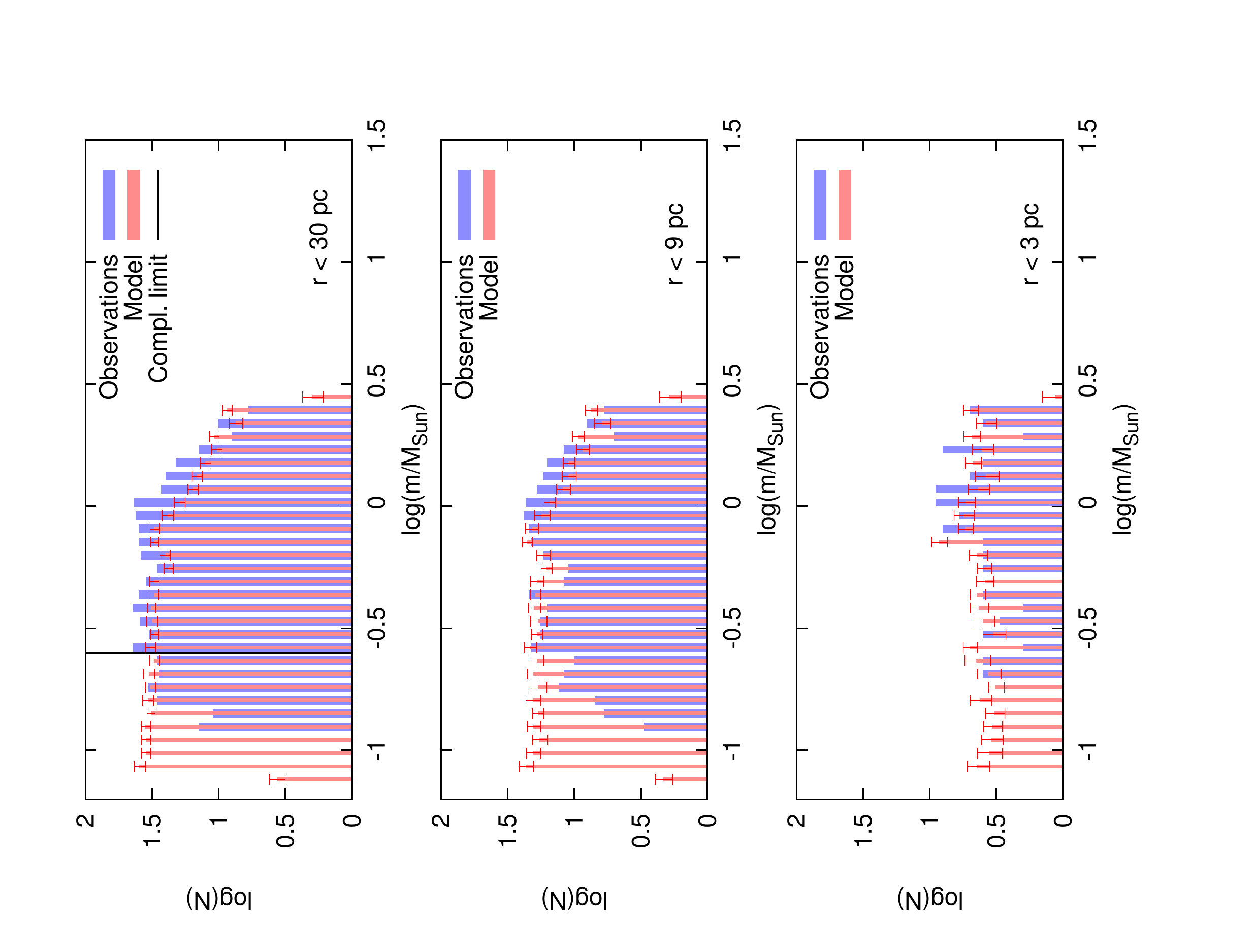} 
\caption{Comparison of the PDMFs between observations and ensemble of models en3000W6.
Top panel: Stars within a radius of 30 pc of the center;
Middle panel: Stars within a radius of 9 pc of the center (Jacobi radius);
Bottom panel: Stars within a radius of 3 pc of the center (core).
Shown is the mean value of $\log(N)$ from the ensemble averaging over 15 runs.
The errors are $1\sigma_m$ standard errors.
The blue bars are derived from the observational data 
with $N=724$ stars and lowest mass $m_l = 0.116 \ M_\odot$, while the 
red bars are derived from the ensemble of models with 
$\langle N_f\rangle=736$ stars and lowest mass $m_l = 0.08 \ M_\odot$.
The onset of incompleteness of the observations is shown as a vertical line
in the top panel.} 
\label{fig:histall}
\end{figure}

\begin{figure}[t]
\centering
\includegraphics[angle=-90,width=0.45\textwidth]{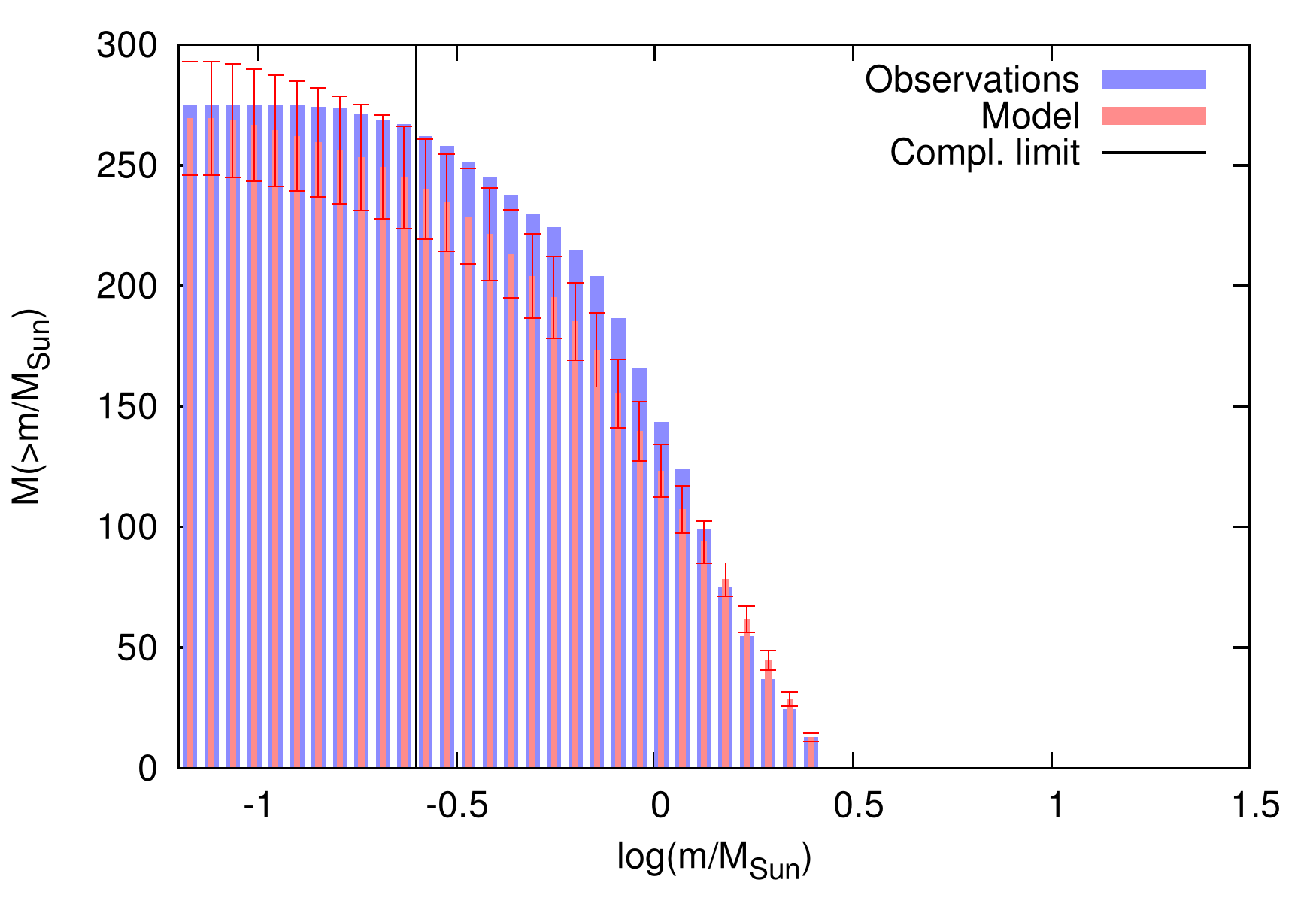} 
\caption{Present-day cumulative mass function for the ensemble en3000W6 for $r < 9$ pc from the cluster
center. Shown is the sum of all masses which are higher than a given mass. The errors
are $1\sigma_m$ standard errors. The completeness limit of the observations 
is shown as a vertical solid line.} 
\label{fig:hist9c}
\end{figure}

Figure \ref{fig:histimf2} shows the average realization of the Kroupa (2001)
IMF of the ensemble en3000W6 with lowest mass $m_l = 0.08 \ M_\odot$ and initial particle number $N_0=3000$. 
Shown is the mean value of $\log(N)$ from the ensemble averaging over 15 runs.
The errors are $2\sigma_m$ standard errors. The analytical
slopes and the transition at $0.5 \ M_\odot$ are also shown as lines.
Note that the depletion in the lowest mass bin is due to the binning.

Figure \ref{fig:histall} shows a comparison of the PDMFs 
between observations and ensemble en3000W6.
The top panel shows the stars within a radius of $r=30$ pc from the center;
the middle panel shows the stars within a radius of $r=9$ pc from the center,
i.e. within the Jacobi radius;
the bottom panel shows the stars within a radius of $r=3$ pc from the center (core).
Shown is the mean value of $\log(N)$ from the ensemble averaging over 15 runs.
The errors are $1\sigma_m$ standard errors.
These two particle numbers $N$ of model and observations
are indeed realized in the top panel. This cannot be seen easily
because of an ``optical illusion'': The deviations between ensemble and observation
carry more weight at high values of $\log(N)$. 
Furthermore, the completeness limit of the observations is around 
$0.25 \, M_{\odot}$ (R\"oser et al. 2011).

The following details can be seen in Fig. \ref{fig:histall}:
In all panels the upper main sequence ($ m > 1 \ M_\odot$)
is overabundant in the observations. The reason may be simply that the IMF of our models
is wrong.
Also, unresolved binaries could be the reason (see Sect. \ref{sec:primordialbinaries}). 
Binaries are not resolved below 3'', i.e. 133 astronomical units
at a distance of 46 pc (R{\"o}ser, priv. comm.).
In the upper two panels, the PDMF of the ensemble is nearly flat at the
low-mass end while the observed PDMF decreases below the completeness
limit of the observations.
A dip in the PDMFs between $m=0.5-0.6 \ M_\odot$ can be seen
in the upper two panels. This dip in the model coincides with the
location of the Wielen dip (Kroupa et al. 1990) which is related to  
the shape of the mass-luminosity relation (Kroupa 2002). The stellar evolution
in our models is different for $m< 0.7\ M_\odot$ and $m> 0.7\ M_\odot$ (Hurley et al. 2000).

Figure \ref{fig:hist9c} shows the cumulative stellar mass function for the ensemble
en3000W6 for stars within a radius of $r=9$ pc of the cluster center,
i.e. within the Jacobi radius. Shown is the sum of all masses that are
higher than a given mass. The histogram includes main sequence stars, giants,
and WDs. Stellar mass BHs are excluded and NSs do not occur. 
Up to the completeness limit of the observations, there is a 
discrepancy of 10 \%, which leaves room for further optimization in future models.

\subsubsection{Stellar evolution}

\begin{table}
\caption{Number counts of stellar types for the best-fitting ensemble of
models (en3000W6).}
\label{tab:en3000W6-stt}
\begin{center}
\begin{tabular}{lllc}
\hline\noalign{\smallskip}
\# & Type & Name &$N \pm \sigma_m$  \\
\noalign{\smallskip}
\hline 
1 & Total & (initial, all stars) & \\
\hline
&0 & Low main sequence $m<0.7 \ M_\odot$ & $2567 \pm 5 $ \\
&1 & Main sequence $m>0.7 \ M_\odot$ &  $432 \pm 5$ \\
\hline
2 & Total & (final, all stars) & \\
\hline
&0 & Low main sequence $m<0.7 \ M_\odot$ & $2567 \pm 5$ \\
&1 & Main sequence $m>0.7 \ M_\odot$ &  $348 \pm 5$ \\
&4 & Core helium burning & $7 \pm 0$ \\
&11 & Carbon-oxygen WDs & $48 \pm 2$  \\
&12 & Oxygen-neon WDs & $7 \pm 1$ \\
&13 & NSs & $13 \pm 1$ \\
&14 & BHs & $5 \pm 0$ \\
\noalign{\smallskip}
\hline
3 & $r < 30$ pc & (final) & \\
\hline
&0 & Low main sequence $m<0.7 \ M_\odot$ & $565 \pm 39$  \\
&1 & Main sequence $m>0.7 \ M_\odot$ & $138 \pm 10$ \\
&4 & Core helium burning & $4 \pm 0$ \\
&11 & Carbon-oxygen WDs & $27 \pm 2$ \\
&14 & BHs & $1 \pm 0$ \\
\hline
4 & $r < 9$ pc & (final, inside Jacobi radius) & \\
\hline
&0 & Low main sequence $m<0.7 \ M_\odot$ & $343 \pm 32$  \\
&1 & Main sequence $m>0.7 \ M_\odot$ & $107 \pm 10$ \\
&4 & Core helium burning & $4 \pm 0$ \\
&11 & Carbon-oxygen WDs & $22 \pm 2$ \\
&14 & BHs & $1 \pm 0$ \\
\hline
5 & $r < 3$ pc & (final, inside core) & \\
\hline
&0 & Low main sequence $m<0.7 \ M_\odot$ & $70 \pm 8$  \\
&1 & Main sequence $m>0.7 \ M_\odot$ & $42 \pm 5$ \\
&4 & Core helium burning & $2 \pm 0$ \\
&11 & Carbon-oxygen WDs &  $9 \pm 1$ \\
&14 & BHs & $1 \pm 0$ \\
\hline
\end{tabular}
\end{center}
\end{table}

\begin{figure}
\includegraphics[angle=-90,width=0.5\textwidth]{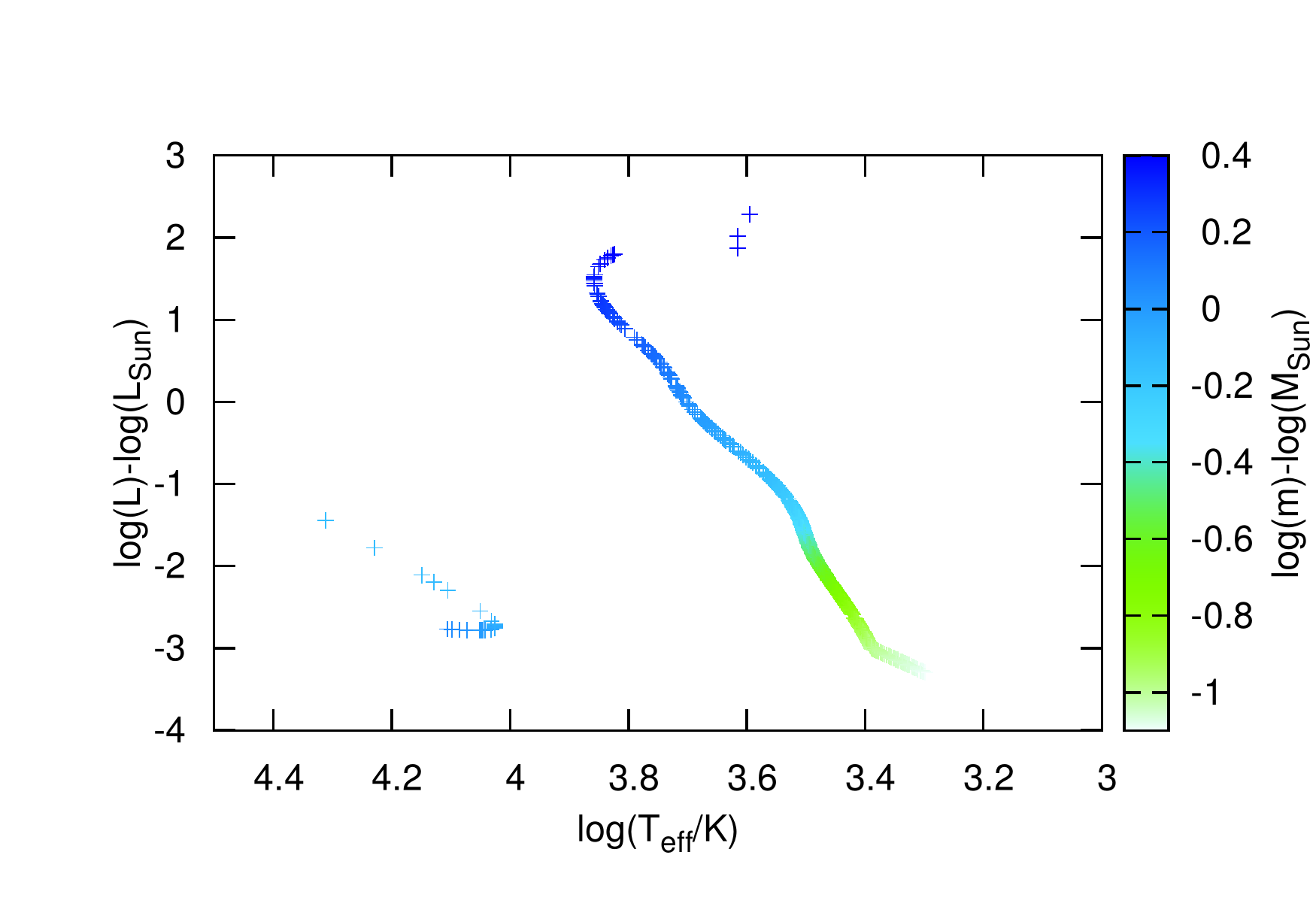} 
\caption{Present-day synthetic Hertzsprung-Russell diagram for one run of the 
evolved model en3000W6. Shown are stars within a radius of $r=30$ pc from the 
cluster center. The final particle number within $r=30$ pc for this run is 
$N_f=836$.} 
\label{fig:hrnew2}
\end{figure}



\begin{figure}[t]
\centering
\includegraphics[angle=-90,width=0.5\textwidth]{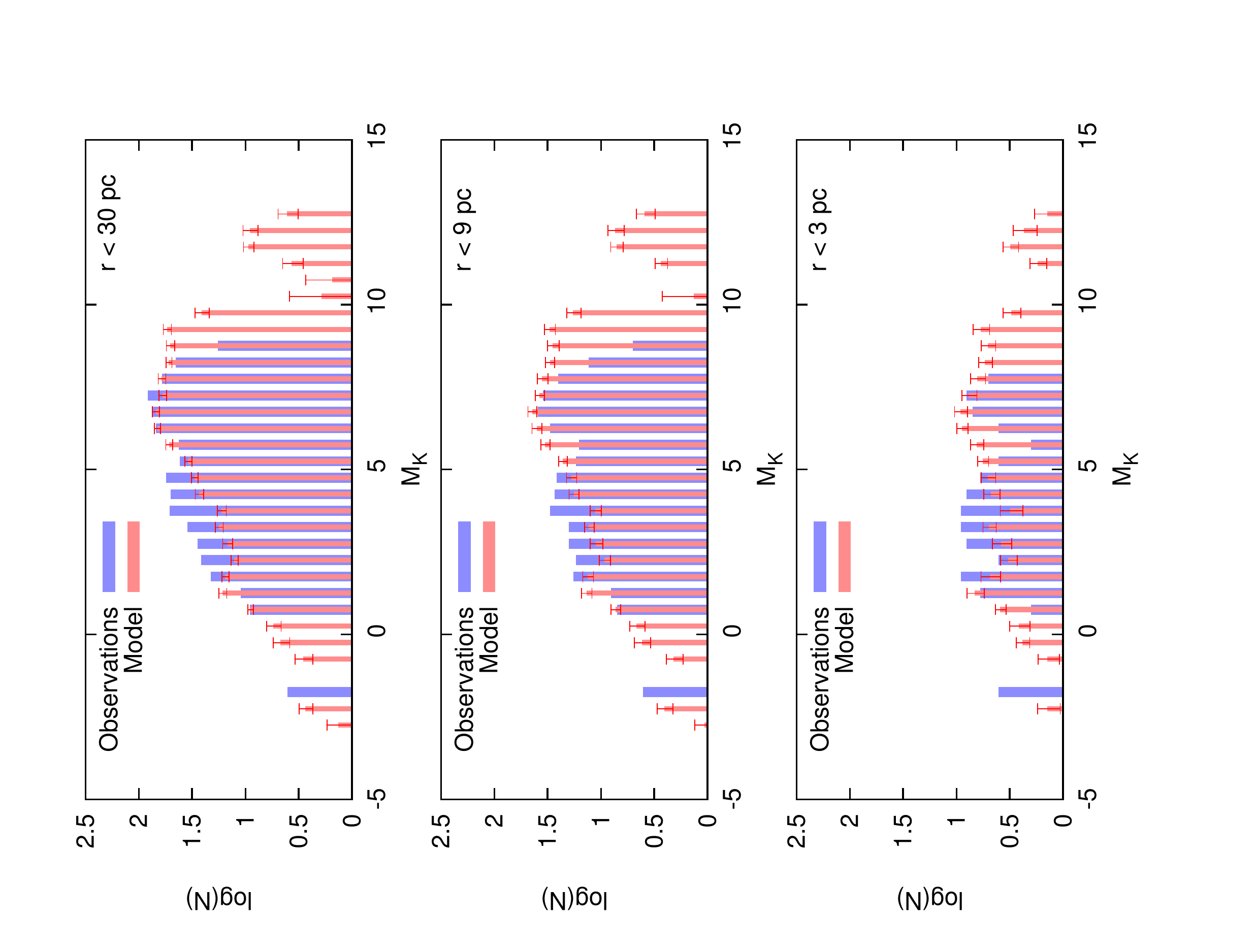} 
\caption{PDLFs in the K band 
for the ensemble en3000W6 and the observations (R{\"o}ser et al. 2011).
For explanations see the text.
Shown is the mean/measured value of $\log(N)$ as a function of the absolute 
$K_s$ (observations)/$K$ (model) band magnitude. 
The errors are $1\sigma_m$ standard errors from the 
ensemble averaging.} 
\label{fig:histallmk}
\end{figure}

Table \ref{tab:en3000W6-stt} shows the number counts of stellar types for 
the best-fitting ensemble of models (en3000W6) for all 3000 stars (initial 
and final), for stars within a radius of 30 pc (final), 9 pc (final inside 
Jacobi radius), and 3 pc (final inside core).
The stellar number counts are averaged over the 15 runs 
of the ensemble. The stellar types are the same 
as in the classification in the beginning of Sect. 4 of 
Hurley et al. (2000). The standard errors from the ensemble-averaging over
15 runs are also given. We find $1\pm 0$ stellar mass BHs within
$r=3$ pc. Almost all stellar mass BHs and all NSs have been kicked out because 
of supernova kicks and are spread out along the cluster orbit. 
We find $27\pm 2$, $22\pm 2$, and $9\pm 1$
carbon-oxygen WDs within $r=30$, $9$, and $3$ pc, respectively.
These numbers can be compared to the observations 
(Schilbach \& R{\"o}ser 2011).



Figure \ref{fig:hrnew2} shows a present-day synthetic Hertzsprung-Russell 
diagram for one run of the evolved ensemble en3000W6. Shown are stars
within a radius of $r=30$ pc from the cluster center. The synthetic
diagram in Madsen (2003) is more suited to direct comparison with
observations since he transforms effective temperatures to $B-V$ colors
using color tables and bolometric magnitudes to V magnitudes. In addition,
he introduces a small scatter in both $V$ and $B-V$. We cannot directly
compare to the observations since the data in $B-V$ are missing for
the data set used in the present study.

Figure \ref{fig:histallmk}
show the comparison of the present-day luminosity functions (PDLFs) 
for the ensemble en3000W6 and the observations by R{\"o}ser et al. 2011.
In the top, middle and bottom panels stars within a radius of 30, 9 (Jacobi radius) and 
3 pc (core) from the cluster center have been used in the statistics.
Shown is the mean value of $\log(N)$ as a function of the 2MASS $K_s$ band magnitude (observations)
and K band magnitude (model)
from the ensemble averaging over 15 runs.
The errors are $1\sigma_m$ standard errors.

To obtain the absolute K band magnitudes from the theoretical
luminosities given by {\sc nbody6}, we applied the compilation of infrared 
bolometric corrections given in Appendix A of Muench (2002) with 
the calibration $m_{\rm bol,\odot}=4.75$. 
The applied conversion formula is

\bea
M_{K} &=& -2.5\log_{10}(L/L_{\odot}) + M_{\rm bol,\odot} - BC_{K}(T_{\rm eff}),
\eea

\noindent
where $M_{K}$, $L/L_{\odot}$, $M_{\rm bol,\odot}$, and $BC_{K}$
are the absolute K band magnitudes, the luminosity in units of the solar luminosity, 
the bolometric magnitude of the Sun, and the K band
bolometric correction, respectively.


In Fig. \ref{fig:histallmk} there are giants in the bin centered at
$M_K = -1.75$ of the observational LF. For these luminous objects it is hard to obtain the
correct observational $M_K$ owing to overexposure (S. Roeser and E. Schilbach,
priv. comm.). On the other hand,
in the theoretical LF the WDs with $\log_{10}(T_{\rm eff})$ = 4.0 - 4.4
populate the faint end ($M_K > 10$). Between $1<M_K<5$ it can be
seen that the observational $M_K$'s are overabundant as compared to
the model $M_K$'s.

\subsubsection{Mass segregation}


\begin{figure}[t]
\centering
\includegraphics[angle=-90,width=0.45\textwidth]{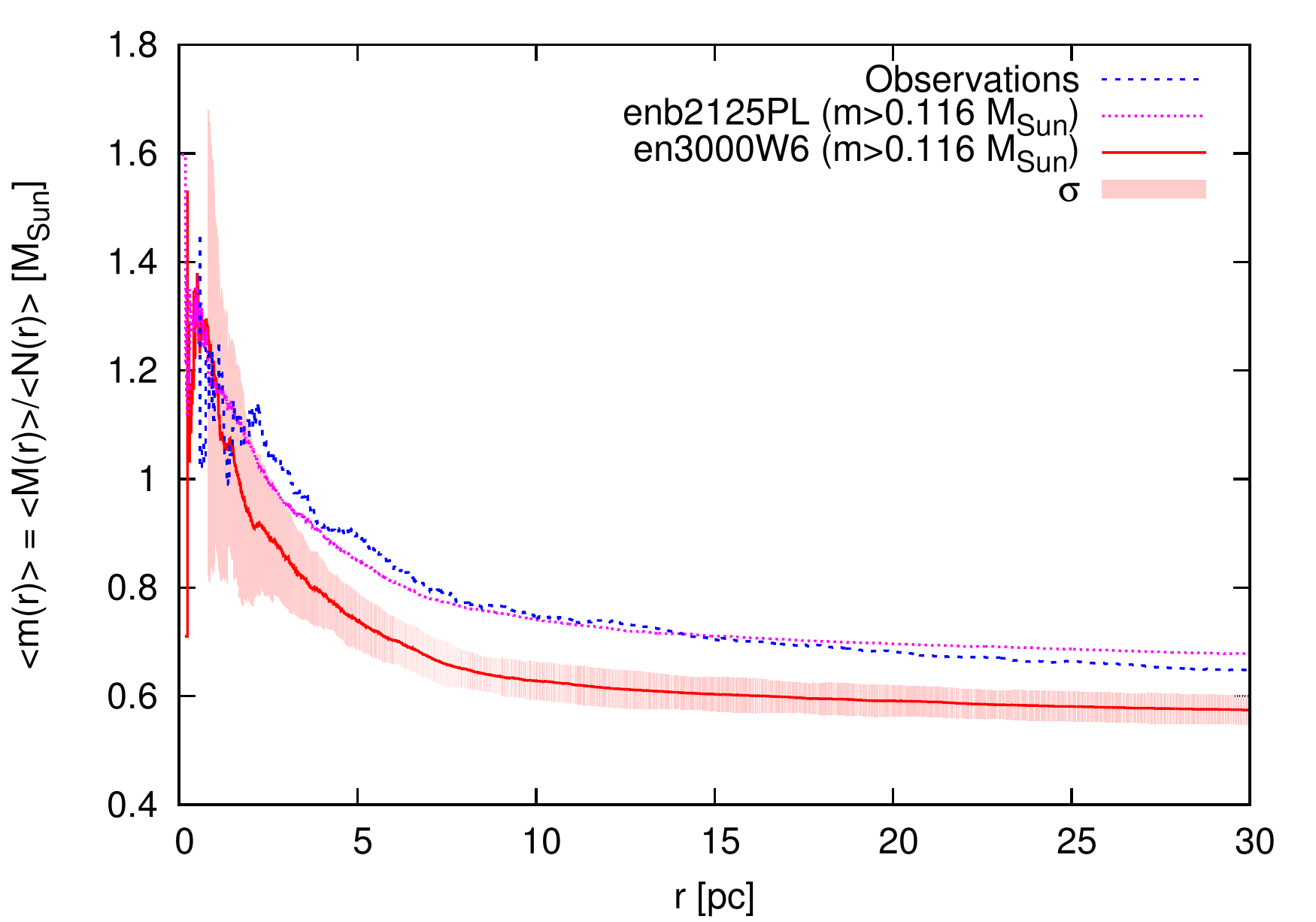} 
\caption{Average cumulative mass for observations and ensembles en3000W6
and enb2125PL.
The average cumulative mass is calculated only for stars with
mass $m\geq 0.116 \ M_\odot$.} 
\label{fig:mav2}
\end{figure}

\begin{figure}[t]
\centering
\includegraphics[angle=180,width=0.45\textwidth]{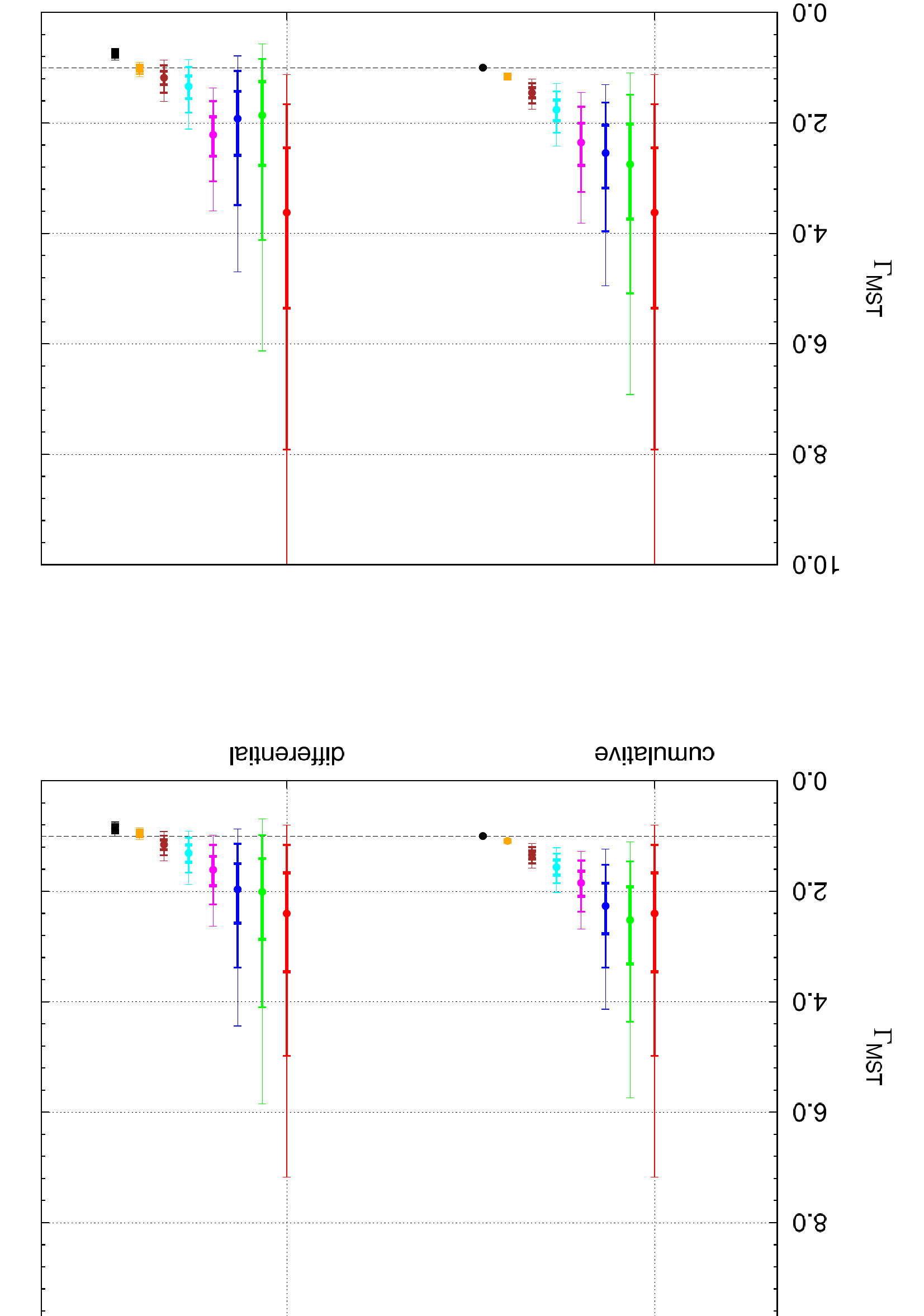} 
\caption{Comparison of present-day mass segregation between models and observations.
Top figure: Analysis of the evolved (for 625 Myrs) ensemble en3000W6. Bottom figure: 
Analysis of the observational
data of R{\"o}ser et al. (2011). Each single plot contains on the left-hand side the ``cumulative'' $\Gamma_{MST}$ (points) for the 5 (red), 10 (green), 20 (blue), 
50 (magenta), 100 (cyan), 200 (brown), 500 (orange), and all most massive stars 
(black). The right-hand side shows the ``differential'' $\Gamma_{MST}$ 
(points) for the 5 (red), 6-10 (green), 11-20 (blue), 21-50 (magenta), 
51-100 (cyan), 101-200 (brown), 201-500 (orange), and 501-all most massive 
stars (black).
The error bars mark the 1, 2, and 3$\sigma$ uncertainties.
A value of one marks the unsegregated state.} 
\label{fig:massseg}
\end{figure}

A first unmistakable sign of mass segregation is the increased mean mass within
the core in Table \ref{tab:en3000W6-par} as compared to the mean mass
within the Jacobi radius.
Figure \ref{fig:mav2} shows the average cumulative mass for observation and ensemble en3000W6. 
The average cumulative mass is calculated only 
for stars with mass $m>0.116 \ M_\odot$.
Also shown is the line for the ensemble enb2125PL with 33\% primordial binaries
(see Section \ref{sec:primordialbinaries} below).
The figure clearly shows the segregation of masses within the Jacobi radius.

Figure \ref{fig:massseg} shows the result of a detailed analysis of mass segregation 
based on the minimum spanning tree (MST) 
method \mbox{${\cal{M}}_{\mathrm{MST}}^{\mathrm{\Gamma}}$}
developed by Olczak, Spurzem \& Henning (2011).
They define a measure of mass segregation $\Gamma_{\rm MST}$,

\bea
\Gamma_{\rm MST} &=& \frac{\gamma_{\rm MST}^{\rm ref}}{\gamma_{\rm MST}^{\rm mass}}, \ \ \ \ \
\Delta\Gamma_{\rm MST} = \Delta\gamma_{\rm MST}^{\rm ref}, \\
\gamma_{\rm MST} &=& \left( \prod_{i=1}^n e_i \right)^{1/n} 
= \exp\left[ \frac{1}{n} \sum_{i=1}^n \ln e_i \right],
\eea

\noindent
where $e_i$ are the lengths of the $n$ MST edges.  The superscript ``ref'' refers 
to a sample of $n + 1$ random stars from the entire population while the superscript ``mass'' refers 
to an equal-sized sample of the most massive stars. 
Note that $\gamma_{\rm MST}$ has the dimension of length while 
$\Gamma_{\rm MST}$ is a dimensionless measure. For $\Delta\Gamma_{\rm MST}$ the geometric standard 
deviation should be used.

The top panel of Figure \ref{fig:massseg} shows 
the analysis of the evolved (for 625 Myrs) ensemble en3000W6, while the bottom 
panel shows an analysis of the  observational data of R{\"o}ser et al. (2011).
Each single plot contains on the left-hand side the ``cumulative"
$\Gamma_{\rm MST}$ (points) for the 5, 10, 20, 50,
100, 200, 500, and all most massive stars (black).
On the right-hand side the ``differential'' $\Gamma_{\rm MST}$ (points) is shown for
the 5, 6-10, 11-20, 21-50, 51-100,
101-200, 201-500, and 501-all most massive stars.
The error bars mark the 1$\sigma$, 2$\sigma$, and 3$\sigma$ uncertainties.
A value of one marks the unsegregated state. The higher the values, the
more segregated the given mass group is (compared to a set of random
groups). The results of this analysis are the following.

\begin{enumerate}
\item The ensemble average of the numerical models resembles the observational 
data very closely. Except for a larger deviation of the fourth bin the
agreement is excellent.
\item Both observations and simulations show a significant degree of
mass segregation, mostly above the $3\sigma$ level for the cumulative
analysis.
\item There is a clear signature of ``inverse mass segregation": $\Gamma_{\rm MST} < 1$
for the last bin in the differential plot (i.e. the $\sim200$ least massive stars).
This demonstrates the removal of the lowest mass members from the inner cluster parts
due to dynamical mass segregation.
\item The signature of mass segregation agrees fairly well with a moderately
($S = 0.3$: cf. \v{S}ubr, Kroupa \& Baumgardt 2008) mass-segregated
model star cluster of 1000 particles as shown in the middle panel of Fig. 4 in
Olczak, Spurzem \& Henning (2011).
\end{enumerate}

\subsubsection{Primordial binaries}

\label{sec:primordialbinaries}

\begin{figure}[t]
\centering
\includegraphics[angle=-90,width=0.45\textwidth]{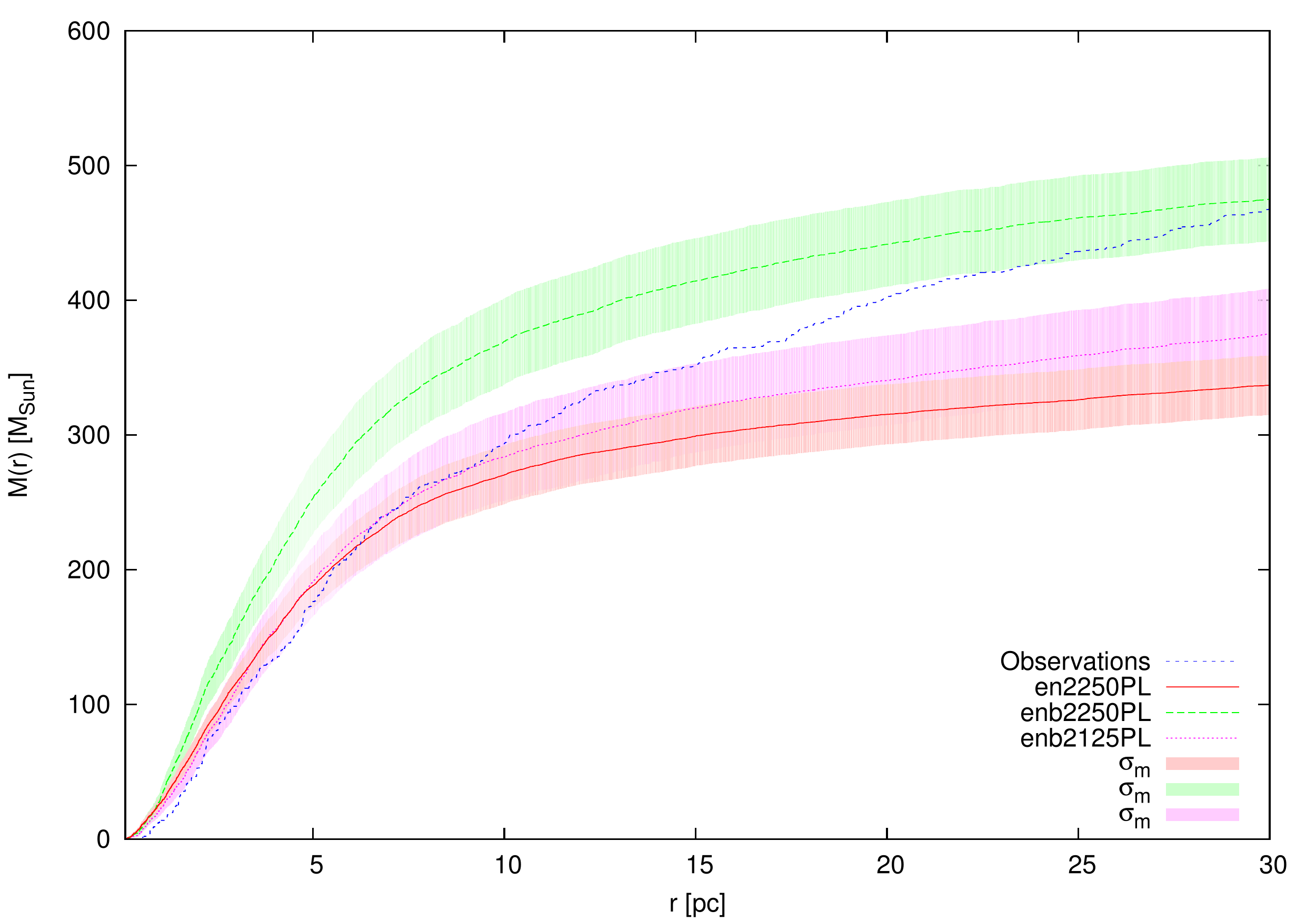} 
\caption{Comparison of the present-day cumulative mass profiles of observations
and models with primordial binaries. 
The thick solid (red) lines are derived from the ensemble en2250PL without
binaries, the dashed (green) line is derived from the ensemble enb2250PL with 33\% 
primordial binaries, and the dotted (magenta) line is derived from the best-fitting ensemble with 
33\% primordial binaries enb2125PL.
The dashed (blue) line represents the observations.
The standard errors are shown as filled areas.
} 
\label{fig:mrmeanpl}
\end{figure}

\begin{figure}[t]
\centering
\includegraphics[angle=-90,width=0.5\textwidth]{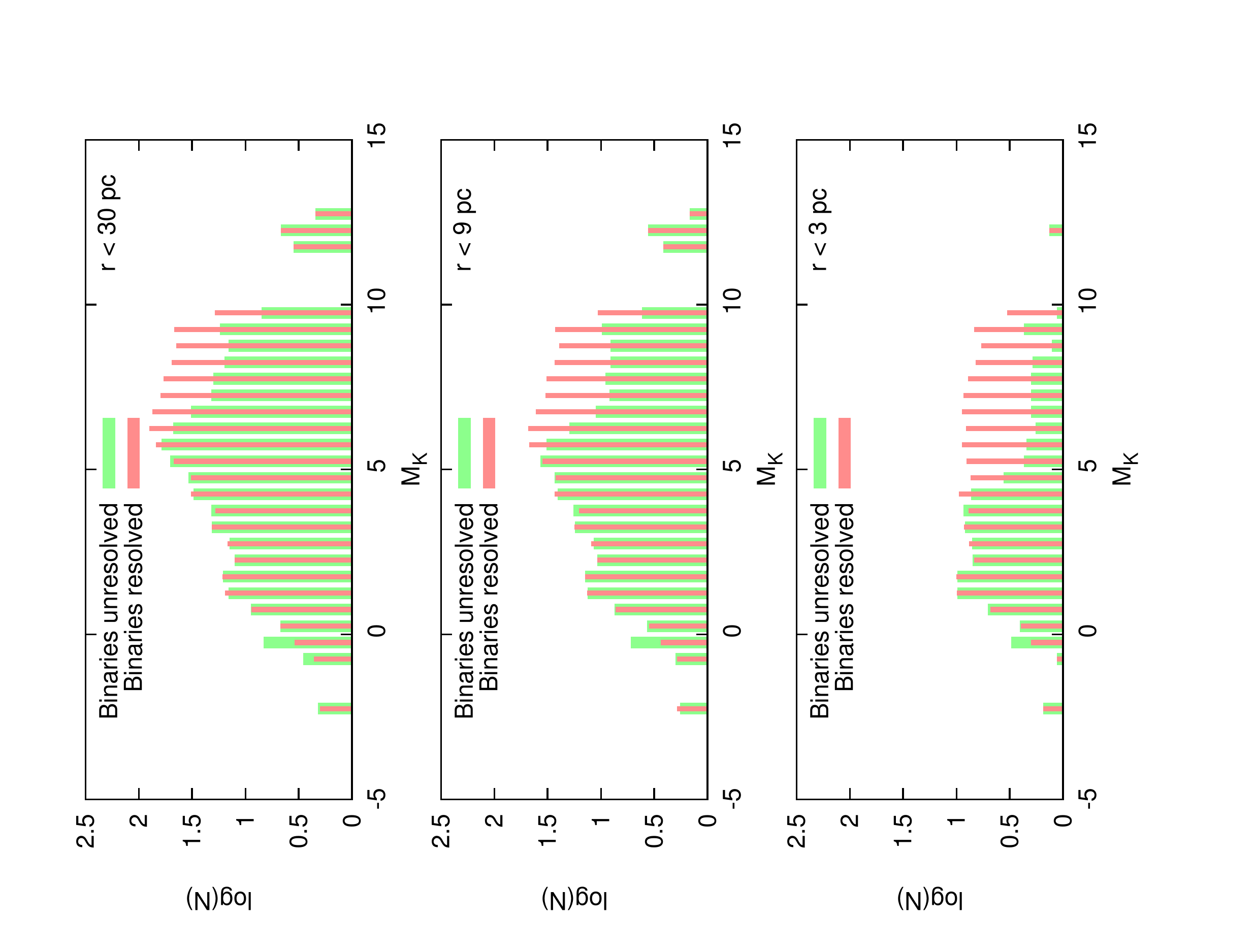} 
\caption{PDLFs in the K band 
for the ensemble enb2125PL where all binaries are either resolved or unresolved. 
Top panel: Stars within a radius of 30 pc of the center;
Middle panel: Stars within a radius of 9 pc of the center (Jacobi radius);
Bottom panel: Stars within a radius of 3 pc of the center (core).
Shown is the mean value of 
$\log(N)$ as a function of the absolute K band magnitude.
} 
\label{fig:histallbmk}
\end{figure}

\begin{figure}[t]
\centering
\includegraphics[angle=-90,width=0.5\textwidth]{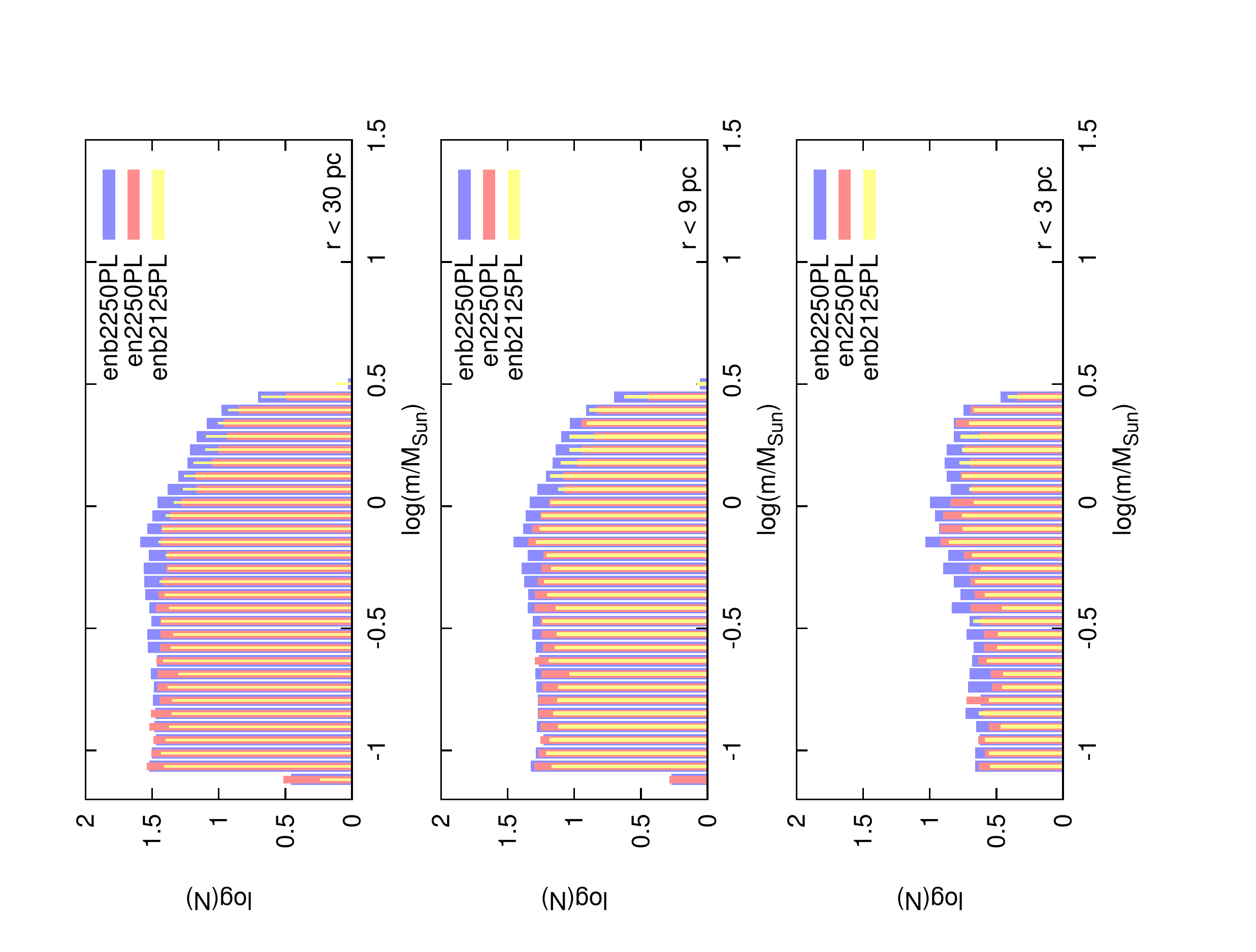} 
\caption{Comparison of the PDMFs between ensembles enb2250PL, en2250PL, 
and enb2125PL.
Top panel: Stars within a radius of 30 pc of the center (Jacobi radius);
Middle panel: Stars within a radius of 9 pc of the center (Jacobi radius);
Bottom panel: Stars within a radius of 3 pc of the center (core).
Shown is the mean value of $\log(N)$ from the ensemble averaging over 15 runs.
The errors are $1\sigma_m$ standard errors.} 
\label{fig:histcmp}
\end{figure}

\begin{table}
\caption{Parameters of the ensemble enb2125PL with 33\% primordial binaries.}
\label{tab:enb2125PL-par}
\begin{center}
\begin{tabular}{lll}
\hline\noalign{\smallskip}
\# & Quantity & Value\\
\noalign{\smallskip}
\hline
1 & Initial model (initial) & \\
\hline
\noalign{\smallskip}
& Particle number $N_0$ & $2125$ \\
& Total mass $\langle M_0 \rangle$ & $1230 \, M_{\odot}$ \\
& Jacobi radius $r_J$ & 14.5 pc \\
\hline
2 & $r<30$ pc (final) & \\
\hline
& Particle number $\langle N_{\rm f}\rangle \pm \sigma_m$ & $619 \pm 58$ \\
& ``Observed'' part. numb. $\langle N_{\rm f} \rangle_{\rm obs} \pm \sigma_m$ & $542 \pm 51$ \\
& Lowest mass $m_l$ & $0.08 \, M_{\odot}$ \\
& Highest mass $m_h \pm \sigma_m$ & $2.86 \pm 0.02 \, M_{\odot}$ \\
& Mean mass $\langle m \rangle$ & $0.606 \, M_{\odot}$  \\ 
& Total mass $M_{\rm f} \pm \sigma_m$ & $375 \pm 35 \, M_{\odot}$ \\
\hline
3 & $r<9$ pc (final, inside Jacobi radius) & \\
\hline
& Particle number $\langle N_{\rm f}\rangle \pm \sigma_m$ & $403 \pm 49$ \\
& ``Observed'' part. numb. $\langle N_{\rm f} \rangle_{\rm obs} \pm \sigma_m$ & $358 \pm 44$ \\
& Mean mass $\langle m \rangle$ & $0.679 \, M_{\odot}$  \\ 
& Total mass $M_{\rm f} \pm \sigma_m$ & $274 \pm 33 \, M_{\odot}$ \\
\hline
4 & $r<3$ pc (final, inside core) & \\
\hline
& Particle number $\langle N_{\rm f}\rangle \pm \sigma_m$ & $127 \pm 21$ \\
& ``Observed'' part. numb. $\langle N_{\rm f} \rangle_{\rm obs} \pm \sigma_m$ & $116 \pm 20$ \\
& Mean mass $\langle m \rangle$ & $0.882 \, M_{\odot}$  \\ 
& Total mass $M_{\rm f} \pm \sigma_m$ & $112 \pm 19 \, M_{\odot}$ \\
\hline
\end{tabular}
\end{center}
\end{table}

\begin{table}
\caption{Number counts of stellar types that are the component of a binary 
for the ensemble enb2125PL.}
\label{tab:enb2125PL-stt}
\begin{center}
\begin{tabular}{lllc}
\hline\noalign{\smallskip}
\# & Type & Name &$N \pm \sigma_m$  \\
\noalign{\smallskip}
\hline
1 & Total & (final, all stars) & \\
\hline
&0 & Low main sequence $m<0.7 \ M_\odot$ & $964 \pm 4$  \\
&1 & Main sequence $m>0.7 \ M_\odot$ & $18 \pm 2$ \\
&11 & Carbon-oxygen WDs & $12 \pm 1$  \\
\hline
2 & $r < 30$ pc & (final) & \\
\hline
&0 & Low main sequence $m<0.7 \ M_\odot$ & $513 \pm 50$  \\
&1 & Main sequence $m>0.7 \ M_\odot$ & $9 \pm 1$ \\
\hline
3 & $r < 9$ pc & (final, inside Jacobi radius) & \\
\hline
&0 & Low main sequence $m<0.7 \ M_\odot$ & $346 \pm 44$  \\
&1 & Main sequence $m>0.7 \ M_\odot$ & $8 \pm 1$ \\
\hline
4 & $r < 3$ pc & (final, inside core) & \\
\hline
&0 & Low main sequence $m<0.7 \ M_\odot$ & $111 \pm 19$  \\
&1 & Main sequence $m>0.7 \ M_\odot$ & $7 \pm 1$ \\
\hline
\end{tabular}
\end{center}
\end{table}

The recent work of Sollima et al. (2009) showed that open clusters may 
contain between $f_b\approx 35 - 70$\% binaries. 
We have run the best-fitting ensemble en2250PL of Plummer models again
with a primordial binary fraction $f_b=33$\% ($f_{sb} = 50$\% 
from Eq. \ref{eq:fb}) in order to investigate the influence of
primordial binaries.  
The ensemble has been called enb2250PL. The third letter ``b'' denotes primordial 
binaries and the following number is the number of stars (i.e. not the number of 
systems).
Since this ensemble does not fit well to the inner observational cumulative mass profile,  
we reduced the initial particle number further and finally found a best-fitting model enb2125PL
(also with 33\% primordial binaries).

Figure \ref{fig:mrmeanpl} shows the comparison of the present-day cumulative mass 
profiles of observations and models with 33\% primordial binaries. 
The standard errors are shown as filled areas.
It can be seen that the profile of the ensemble enb2250PL is steeper than that of
the ensemble en2250PL at small radii.
The ratio of half-mass to Jacobi radius for the model enb2125PL is given by
$r_h/r_J = 0.18 \pm 0.01$.
The field star contamination for the model enb2125PL is $|M_{\rm model}- M_{\rm obs}|/M_{\rm obs}=$ 
$12\% \ (3 \ \rm pc)$,  $0\% \ (9 \ \rm pc)$, $13\% \ (18 \ \rm pc)$, and $20\% \ (30 \ \rm pc)$.



Figure \ref{fig:histallbmk} shows the PDLFs in the K band 
for the ensemble enb2125PL where all binaries are either resolved or unresolved. 
For the unresolved binaries we transformed for each star the bolometric luminosities 
given by {\sc nbody6} to absolute K band magnitudes using the bolometric corrections of 
Muench (2002) and then back to K band luminosities
using the absolute K-band magnitude of the Sun $M_{K,\odot} = 3.33$. 
Then we added the K band luminosities 
of both binary components and transformed back to absolute K band magnitudes,
which are shown in Fig. \ref{fig:histallbmk}. With the present parameters of the
distribution of binaries, unresolved binaries produce a difference at the 
faint end of the LF. However, our statistics of the binary parameters
may be not fully realistic.

Figure \ref{fig:histcmp} shows a comparison of the PDMFs between ensembles 
enb2250PL, en2250PL, and enb2125PL. It can be seen that the ensembles
en2250PL and enb2125PL are depleted of high-mass stars as compared 
to the ensemble enb2250PL for $r < 30$ pc and $r < 9$ pc. Furthermore, the comparison of
ensembles en2250PL and enb2125PL shows that the ensemble enb2125PL
is depleted of low-mass stars compared to the ensemble en2250PL,
while high-mass stars are overabundant in the ensemble enb2125PL
as compared with the ensemble en2250PL.

The initial and final parameters of the ensemble enb2125PL
are given in Table \ref{tab:enb2125PL-par}. 
We note that (1) the final parameters are obtained after 625 Myrs of evolution, i.e. at the present time,
(2) the final ``observed'' particle number is the number of stars with $m > 0.116 \ M_{\odot}$ and that
(3) the highest mass is obtained by excluding BHs.

Table \ref{tab:enb2125PL-stt} shows the number counts of stellar types 
that are the component of a binary for the ensemble enb2125PL for all 
2125 stars (final), for stars within a radius of 30 pc (final), 9 pc 
(final inside Jacobi radius), and 3 pc (final inside core). The stellar 
number counts are averaged over the 15 runs of the ensemble. 
The stellar types are the same as in the classification 
in the beginning of Sect. 4 of Hurley et al. (2000).
There are no BHs or NSs in binaries. Also, there 
are no WDs in binaries within 30 pc  radius of
the cluster center. They are all kicked out by binary kicks.
Since WDs in binaries are observed in the Hyades (Schilbach \& R{\"o}ser 2011) 
they may not be subject to strong kicks.

\subsection{Future evolution}

The future evolution of the ensemble en3000W6 is shown in the inset of 
Figure \ref{fig:ntall}. There the times when the cluster has reached
20, 10 and 5\% of its initial particle number are shown. The times are $t_{20} = 695$ Myrs,
$t_{10} = 875$ Myrs, and $t_{5} = 1029$ Myrs.  The last time may be taken
as an approximate dissolution time.

\label{sec:futureevolution}

\section{Discussion}

\label{sec:discussion}

The present-day properties of the
Hyades can be reproduced well by a standard King or Plummer initial model when choosing appropriate
initial conditions. We have found a model (en3000W6), which imitates the cumulative 
mass profile within the Jacobi radius of the observed Hyades spatial distribution very well. 
Also, the derived PDMF and the K band LF show a good agreement
between observations and model.

From our models we can make the following statements about the Hyades:

\begin{itemize}
\item The tidal tails of the Hyades have a length of $\approx 800$ pc
today if they are not destroyed by passing giant molecular clouds,
spiral arm passages, disk shocking or other effects.
\item Roche-lobe filling $W_0=6 - 9$ single-star King initial models (King 1966) with the initial 
particle number $N_0 = 3000$ provide a good fit to the present-day inner cumulative 
mass profile. 
\item The best-fit, single-star, Roche-lobe filling King model has an initial mass of 
$M_0=1721 \ M_\odot$, an initial
Jacobi radius of $r_J=16.2$ pc, and an average
mass loss rate of $dM/dt=2.2 \ M_\odot/\mathrm{Myr}$.
\item A reasonably good fit is obtained with a Plummer model with 33\% primordial binaries
and a ratio of half-mass to Jacobi radius of $r_h/r_J = 0.18$. 
Here the initial particle number is $N_0 = 2125$ and mass are is $M_0 = 1230 \ M_\odot$. 
The average mass loss rate is $dM/dt= 1.4 \ M_\odot/\mathrm{Myr}$.
The observed average cumulative mass is reproduced very well.
\item Mass segregation is clearly detected in both the observations and models of the Hyades
cluster.
\item The Hyades contain only $1 \pm 0$ stellar mass black holes. The
probability is very high that nearly all NSs and BHs are 
kicked out by supernova kicks.
\item The number of WDs critically depends on the kick velocity
that is adopted for WD kicks. This kick velocity has not yet been well constrained.
Assuming a mean kick velocity of $\approx 8$ km/s (comparable to the escape
velocity from the cluster center), we obtain $27 \pm 2$,  $22 \pm 2$, and $9 \pm 1$ carbon-oxygen
WDs within a radius $r<30$, $r<9$ (Jacobi radius), and $r<3$ pc (core) from the cluster center,
respectively. For the comparison to the observations we refer to
the article by Schilbach \& R{\"o}ser (2011). 
We do not find WDs in binaries within $r < 30$ pc in our model. They are all kicked out
in the form of binary kicks. However, some WDs in binaries are observed in the Hyades
(Schilbach \& R{\"o}ser 2011).  This suggests that WDs in binary systems are not subject to
strong kicks.
\item The including of 33\%  primordial binaries reduces the initial number of Hyades 
stars $N_0$ by $\approx 5\%$ as compared to the model without primordial binaries.
\item Under the assumption that the single-star ensemble en3000W6 is a good 
approximation of reality, we find that the Hyades will 
be dissolved except for 5\% of its initial number of stars in $\approx 400$ Myrs from now.
\end{itemize}

The degeneracy of good-fitting models can be quite high due to the large 
dimension of the parameter space, which is spanned by the initial particle number 
$N_0$, the King parameter $W_0$, and two extra dimensions: the primordial binary 
fraction $f_b$ and the Roche-lobe filling factor $r_{99\%}/r_J$. 
In this work we did not explore the latter two degrees of freedom  in detail.
More simulations with different Roche-lobe filling
factors and primordial binary fractions are required to explore this degeneracy in more detail.

\section{Acknowledgements}

This work was supported by the German Research Foundation (DFG) under 
the Collaborative Research Center SFB 881 (``The Milky Way System'') 
at the University of Heidelberg.

Simulations were performed on the GRACE supercomputer
(grants I/80 041-043 and I/84 678-680 of the Volkswagen Foundation
and 823.219-439/30 and /36 of the Ministry of Science, Research and
the Arts of Baden-W{\"u}rttemberg).

CO appreciates funding by the German Research Foundation (DFG), grant
OL~350/1-1.

PB acknowledges the special support by the NAS Ukraine under
the Main Astronomical Observatory GRAPE/GPU/GRID computing
cluster project. PB's studies are also partially supported by the
program Cosmomicrophysics of NAS Ukraine.

The authors would like to thank Siegfried R{\"o}ser and Elena Schilbach for
providing their observational data and the reading of the manuscript, 
Keigo Nitadori and Sverre Aarseth
for making {\sc nbody6gpu} available and Jarrod Hurley for developing
of the stellar evolution routines in {\sc nbody6}.

Furthermore, useful conversations with Siegfried R{\"o}ser, Elena Schilbach 
and Sverre Aarseth are also acknowledged.

AE would like to thank his father, Reinhard Ernst, for a discussion.

\bigskip

{\bf Note added in proof:} The first author forgot to remove the white dwarfs from the
simulation data. However, removal of the white dwarfs does not resolve the discrepancy between
observations and model en3000W6, e.g. in Figures 8 and 11.

\appendix

\section{nbody6tidgpu}

\label{sec:nbody6tidgpu}

The program {\sc nbody6tidgpu} is a new direct $N$-body code
for integrating the $N$-body problem in an analytic
background potential using Graphics Processing Units (GPUs). 
The program is essentially based on {\sc nbody6} by Aarseth 
(Aarseth 1999, 2003).
{\sc nbody6tidgpu} is not parallelized with MPI routines but instead runs on 
multicore processors (via OpenMP) with one or two GPU(s). For this
purpose, special routines in CUDA have been added by Nitadori in collaboration
with Aarseth. The variant of {\sc nbody6} with GPU
support is termed {\sc nbody6gpu} (Nitadori \& Aarseth 2011).

Implementing different galactic tidal fields into {\sc nbody6gpu} is 
the contribution of some of the present authors and the special feature 
of {\sc nbody6tidgpu}. This implementation is based on the implementation 
in the (parallel) code {\sc nbody6gc} which is described in detail in 
Ernst, Just \& Spurzem (2009) and Ernst (2009). Several analytic models 
for the background potential are part of the code, for example,

\begin{itemize}
\item the three-component Plummer-Kuzmin model of the Milky Way described 
in Section \ref{sec:numericalmethod} (Miyamoto \& Nagai 1975)
\item power-law models with and without a central black hole
\item Dehnen models (Dehnen 1993, Tremaine et al. 1994)
\item Plummer model (Plummer 1911).
\end{itemize}

In addition to the equations of motion of the $N$-body problem, 
the program solves the equations of motion of the star cluster orbit 
around the Galactic center with an eighth-order composition scheme 
(Yoshida 1990; for the coefficients see McLachlan 1995). The tidal 
force of the background potential acts on all particles in the 
$N$-body system and is added as a perturbation to the KS 
regularization (Kustaanheime \& Stiefel 1965) of {\sc nbody6}. 

The program is able to handle the integration of star cluster
orbits with almost all eccentricities. For nearly radial orbits a
time transformation can be switched on to guarantee an exact orbit 
integration during the pericenter passages. The leapfrog-like symplectic 
integration schemes allow for adaptive time steps if one applies a 
Sundman transformation to the time (Sundman 1912, Mikkola \& Tanikawa 1999, 
Preto \& Tremaine 1999, Mikkola \& Aarseth 2002).

It is also possible to switch on dynamical friction with different $\chi$ functions 
and different realizations of the Coulomb logarithm, i.e. fixed or variable
according to Just \& Pe\~narrubia (2005). 
Since the symplectic composition schemes are by construction suited to
Hamiltonian systems, the dissipative dynamical friction force requires 
special attention. It is implemented in {\sc nbody6tidgpu} with an iterative 
implicit midpoint method (see, e.g., Mikkola \& Aarseth 2002).

For the current study, the three-component Plummer-Kuzmin model described 
in Section \ref{sec:numericalmethod} is used as a background potential.
The acceleration for the Plummer-Kuzmin model is given by the expression

\be
\left( \begin{array}{c}
A_x \\
A_y \\
A_z \\
\end{array} \right) = - \frac{ GM }{ \left( R^2 + (a + \sqrt{b^2 + z^2} )^2 \right)^{3/2}}\cdot \left( \begin{array}{c}  
x \\
y \\
z\cdot \frac{a+\sqrt{b^2+z^2}}{\sqrt{b^2+z^2}} \\
\end{array} \right) \label{eq:eq-gal}
\ee

\noindent
For the tidal correction to the equations of motion of the $N$-body problem, 
the time derivative of acceleration (jerk) for the analytic background
potential is needed to guarantee energy conservation
and exactness of the orbit integration on a high level.
The jerk is given by lengthy 
expressions that depend also on the velocity components $V_x, V_y$, and 
$V_z$ and that we do not reproduce here. However, they are
also implemented in {\sc nbody6tidgpu}.

\section{Bad fits}

\label{sec:badfits}

\begin{figure}[t]
\centering
\includegraphics[angle=-90,width=0.5\textwidth]{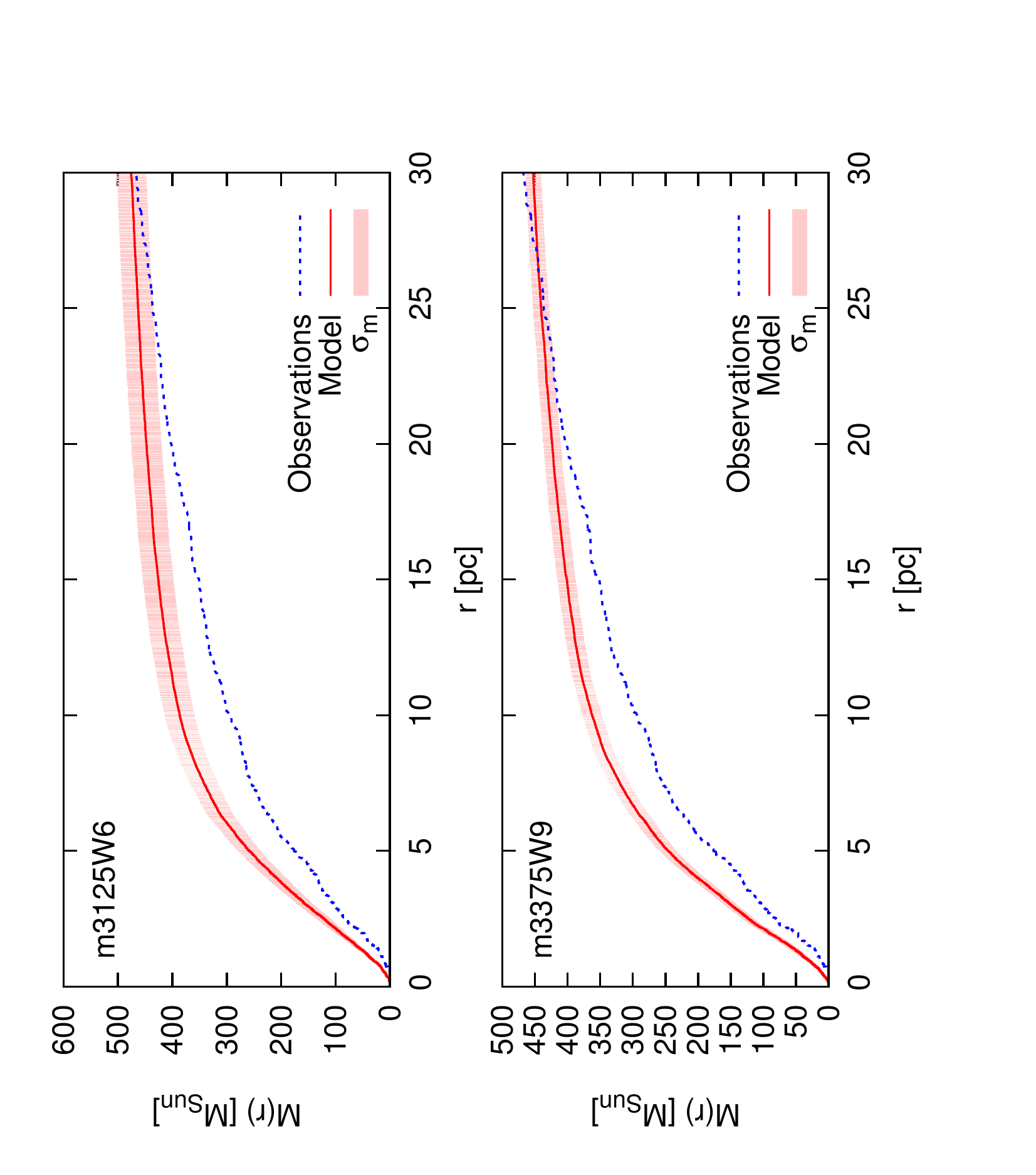} 
\caption{Comparison of the cumulative mass profiles of observations
and ensemble of models. The thick solid (red) lines in the middle for the model  
represent the mean that is ensemble-averaged over 15 runs. 
The standard error of the mean for the ensemble is shown as a filled area.
The dashed (blue) line is calculated from the observational data.
Top: Ensemble of models en3125W6.
Bottom: Ensemble of models en3375W9.}
\label{fig:mrmean4b}
\end{figure}

Figure \ref{fig:mrmean4b} shows the two best fits for the total mass
of Hyades members within 30 pc from the cluster center for the models
within the second parameter grid of Section \ref{sec:parameterspace},
i.e. the grid with $\Delta N = 125$. 
The top panel shows the ensemble of models en3125W6 $(N_0 = 3125, W_0 = 6)$ while
the bottom panel shows the ensemble of models en3375W9 $(N_0 = 3375, W_0 = 9)$.
It can be seen that the agreement between observations and ensemble of models is
not even consistent within $2\sigma_m$ for the largest part of the radial range.

\end{document}